\documentclass[fleqn,usenatbib]{mnras}
\usepackage[T1]{fontenc}
\DeclareRobustCommand{\VAN}[3]{#2}
\let\VANthebibliography\thebibliography
\def\thebibliography{\DeclareRobustCommand{\VAN}[3]{##3}\VANthebibliography}

\usepackage{graphicx}	% Including figure files
\usepackage{amsmath}	% Advanced maths commands
\usepackage{amssymb}	% Extra maths symbols
\usepackage{color}
\usepackage{comment}
\usepackage{threeparttable}
\usepackage{subfig}

\title[Magnetohydrodynamic convection in accretion discs]{Magnetohydrodynamic convection in accretion discs}

\author[L.E. Held et al.]{
Loren E. Held,$^{1,2}$\thanks{E-mail: loren.held@aei.mpg.de (LEH)}
Henrik N. Latter$^{1}$
\\
$^{1}$Department of Applied Mathematics and Theoretical Physics, University of Cambridge, Centre for Mathematical Sciences,\\
 Wilberforce Road, Cambridge CB3 0WA, UK\\
 $^{2}$Max Planck Institute for Gravitational Physics (Albert Einstein Institute), Am M{\"u}hlenberg 1, Potsdam 14476, Germany
}

\date{Accepted XXX. Received YYY; in original form ZZZ}

\pubyear{2020}

\begin{document}
\label{firstpage}
\pagerange{\pageref{firstpage}--\pageref{lastpage}}
\maketitle

\begin{abstract}
Convection has been discussed in the field of accretion discs for several decades, both as a means of angular momentum transport and also because of its role in controlling discs' vertical structure via heat transport. If the gas is sufficiently ionized and threaded by a weak magnetic field, convection might interact in non-trivial ways with the magnetorotational instability (MRI). Recently, vertically stratified local simulations of the MRI have reported considerable variation in the angular momentum transport, as measured by the stress to thermal pressure ratio $\alpha$, when convection is thought to be present. Although MRI turbulence can act as a heat source for convection, it is not clear how the instabilities will interact dynamically. Here we aim to investigate the interplay between the two instabilities in controlled numerical experiments, and thus isolate the generic features of their interaction. We perform vertically stratified, 3D MHD shearing box simulations with a perfect gas equation of state with the conservative, finite-volume code \textsc{PLUTO}. We find two characteristic outcomes of the interaction between the two instabilities: straight MRI and MRI/convective cycles, with the latter exhibiting alternating phases of convection-dominated flow (during which the turbulent transport is weak) and MRI-dominated flow. During the latter phase we find that $\alpha$ is enhanced by nearly an order of magnitude, reaching peak values of $\sim 0.08$. In addition, we find that convection in the non-linear phase takes the form of large-scale and oscillatory convective cells. Convection can also help the MRI persist to lower Rm than it would otherwise do. Finally we discuss how our results help interpret simulations of Dwarf Novae.
\end{abstract}


\section{Introduction}
\label{INTRO}
The prevalence and role of convection perpendicular to the plane of accretion discs has been discussed for nearly as long as the modern field of discs itself has been around \citep{armitage2011,klahr2006, balbus1998instability, pringle1981}. Heating of the disc, by accretion or otherwise, and the subsequent vertical transport of heat sets the vertical structure of the disc. Whether this transport is dominated by radiation, or is facilitated by a mixture of radiation and convection, depends, in general, on the balance of vertical entropy, compositional, and radiative gradients: if the disc is optically thick, the combination of heating \textit{and} its effective trapping inside the disc leads to an unstable entropy gradient that is conducive to thermal convection, while partial ionisation of material can set up unstable compositional gradients. Indeed, simple 1D vertical disc structure models in which viscous heating is prescribed through an effective $\alpha$ viscosity are known to be convectively unstable for many realistic opacities (e.g. Hydrogen-ionisation, melting of ice grains, evaporation of metal grains, and even Thompson opacity provided there is sufficient radiation pressure), demonstrating that convection may be a generic feature of many discs \citep{meyer1982, lin1980structure,bisnovatyi1977disk,pollack1985}.

While the aforementioned models do not specify the exact mechanism for the effective viscosity, sufficiently ionized discs threaded by a weak magnetic field are unstable to the magnetorotational instability (MRI), which is thought to supply the mechanism of angular momentum transport and associated viscous dissipation. But convection and the MRI share similar dynamical scales and thus the latter is unlikely to appear as a mere viscous heat source; the interaction between the MRI and convection will be far more complicated. Indeed, there is numerical evidence that the two instabilities combine in striking ways; under certain circumstances they enhance angular momentum transport by as much as an order of magnitude \citep{hirose2014,hirose2015,scepi2018a,coleman2018}. Our aim is to explore the range of this interplay through controlled numerical experiments. Before reporting our results, however, we first discuss the prevalence of convection in various classes of disc in Section \ref{INTRO_Applications}, and summarize some of the key outstanding problems in Section \ref{INTRO_KeyQuestions}.

\subsection{Applications}
\label{INTRO_Applications}
Convection has featured in the disc instability model explaining the outbursts of dwarf novae, in radiation-pressure dominated discs relevant to low-mass x-ray binaries (and potentially AGN), and has been also been hypothesized to exist in protoplanetary discs at various radii and times in their evolution. We summarize these applications below. 

\subsubsection{Convection in dwarf novae}
Convection has long featured in theoretical models of \textit{dwarf novae} (DNe), binary systems in which a white dwarf accretes from a low mass companion via Roche lobe overflow (\cite{warner1995, hellier2001}). A defining observational feature of dwarf novae is the presence of semi-regular outbursts, which are thought to be the result of an instability in the accretion disc surrounding the white dwarf. This `disc instability' model relies on the system supporting equilibria following an `S-curve' in the plane of temperature against surface density. The disc then cycles between cool, dim, inefficiently accreting `low'-states (quiescence) on the lower branch of the S, and hot, bright, efficiently accreting `high'-states (outburst) on the upper branch \citep{lasota2001}. There was initially some debate over the physics that leads to the hysteresis, and for a period it was thought that convection was the culprit \citep{meyer1981, meyer1982,smak1982a,cannizzo1982convective}. Ultimately, it was demonstrated that ionisation of hydrogen at temperatures around $10^4$\,K provided the underlying cause. This ionisation transition results in an opacity with a strong temperature dependence, allowing for the multiple thermal equilibria required by theory \citep{hoshi1979,faulkner1983evolution}. 

Although it was demonstrated that convection was not responsible for the cyclical outbursts in dwarf novae, these early calculations indicated that convection was a natural \textit{consequence} of the vertical variation in opacity (through ionisation) needed to produce these outbursts. Thus convection is expected to be a feature of `warm' thermally stable states around $10^4$ K, and may also be enhanced by the unstable composition gradients brought about by the self-same ionisation. Indeed, a number of recent local MHD simulations have claimed to find convection  both in DNe conditions \citep{hirose2014,scepi2018a} as well as in AM CVn conditions \citep{coleman2018}. We engage with this literature in more detail in Sections \ref{INTRO_KeyQuestions} and \ref{DISCUSSION_ApplicationsToDwarfNovae}.

\subsubsection{Convection in LMXBs and AGN}
Convection might be also be present in low mass X-ray binaries (LMBXs), systems in which a black hole (BH) or neutron star (NS) primary accretes via Roche Lobe overflow from a low mass secondary. Like dwarf novae, LMXBs are observed to undergo repeated outbursts. The outburst phase in LMXBs, however, is much more complicated than in dwarf novae, and comprises several distinct spectral states. `Super soft' and `high/SPL' states are characterized by very high accretion rates and are modelled by a geometrically thin, optically thick disc that extends all the way down to the central object \citep{done2007}. The disc in these states is likely radiation pressure dominated and hence convectively unstable (and indeed thermally unstable), at least according to simple alpha disc models \citep{bisnovatyi1977,tayler1980,pietrini2000,blaessocrates2001}.

The disc morphology in the `low/hard' and quiescent states is more involved. The relatively cool, geometrically thin, optically thick outer regions of these discs are similar to those of dwarf novae. The inner regions, however, differ substantially: they are extremely hot, optically thin, and geometrically thick (bearing greater similarity to tori than to thin discs).  These thick structures are also believed to manifest in low-luminosity AGN \citep{fabianrees1995,yuan2014}. Several theoretical models have been developed to describe these hot flows. One such class, the advection dominated accretion flows (ADAFs) \citep{narayanyi1994}, were shown to be \textit{very} convectively unstable in the plane of the disc, so much so that the stabilizing influence of rotation is overcome. ADAFs are likely to be even more convectively unstable \textit{perpendicular} to the disc (where the Solberg-Hoiland criterion reduces to the less stringent Schwarzchild criterion).  Various extensions of, and modifications to, these models have been introduced, including advection dominated inflow outflow solutions (ADIOS) \citep{blandfordbegelman1999}, and convection dominated accretion flows (CDAFs)  \citep{quataertgruzinov2000}. The latter, in particular, posit that convection has forced the system to a marginally unstable state (according to the Solberg-Hoiland criterion), a condition that then completely determines the resulting flow's structure. However, global MHD simulations that supply the turbulent dissipation self-consistently (via the MRI) do not map to these solutions in any obvious way (e.g. \cite{stonepringle2001,hawley2001magnetohydrodynamic,hawleybalbus2002}).

\subsubsection{Convection in Protoplanetary discs}
The presence of thermal convection in cooler protoplanetary discs is less well established, mainly on account of their weak internal heating. Beyond a handful of AU, irradiation of the disc by the central star, rather than accretion, dominates the thermodynamics, resulting in a roughly isothermal vertical structure at larger radii, or cooler midplanes and hotter surfaces at smaller to intermediate radii  \citep[e.g.][]{DAlessio1998}. Both are situations not conducive to convection. This point was not appreciated in early work that put convection front and centre in the disc's evolution (e.g. \cite{rudenlin1986,cabot1987a,cabot1987b,rudenpollack1991}.

Another issue is the provenance of the midplane heating, as the driver of angular momentum transport in these objects has not been unambiguously identified (see \cite{turner2014}). Discs around young stars are typically cold, and insufficiently ionized to support the MRI outside some critical radius $\lesssim 1\,$AU. Nevertheless, there are various heat sources that can, in principle, drive convection at certain times and locations. Very young systems not yet in quasi-equilibrium may be unstable on account of their vertical contraction and opacity regime \citep{lin1980structure}. Later in their evolution a potential heat source may be the dissipation of strong shocks launched by a high mass planet \citep{lyra2016shocks}, or dissipation of magnetic fields facilitated through ambipolar or ohmic diffusion in mid-plane current sheets \citep{bethune2017, bethune2020}. Generally, the optically thick and MRI-unstable inner radii ($< 1$ AU) should also generate adverse entropy profiles in $z$ \citep[e.g.][]{hirose2015}, and during outburst this active region may extend to significantly larger radii ($5-10$ AU) (e.g. \cite{armitage2001,zhu2009nonsteady,zhu2010long,kadam2020outbursts}.

\subsection{Key questions}
\label{INTRO_KeyQuestions}
If convection is present in discs, a question that naturally arises is: can convection drive accretion? \cite{cameron1969} and \cite{paczynski1976} were perhaps the first to suggest that an effective viscosity due to convection might drive accretion in protosolar and dwarf novae discs (respectively). This led to the concept of self-sustaining convection in discs, by which convective motions supply a turbulent viscosity that can thermalise sufficient energy from the background shear to maintain the self-same convection (cf. several mixing length models: \cite{vila1978,lin1980structure,smak1982a,rudenlin1986,cannizzocameron1988}).  Though an attractive idea, hydrodynamic simulations conducted by \cite{stone1996angular} and \cite{heldlatter2018} have failed to observe self-sustaining convection.

A related question is whether convection even transports angular momentum with the correct sign to drive accretion. Early linear studies \citep{ryu1992convective,lin1993nonaxisymmetric} and numerical simulations \citep{,cabot1996numerical,stone1996angular} made conflicting claims as to the sign of the angular momentum transport. The most recent results from 3D hydrodynamic shearing box simulations indicate that hydrodynamic convection \textit{can} transport angular momentum outwards \citep{lesur2010angular,heldlatter2018}, with \cite{heldlatter2018} finding that the negativity of the sign of $\alpha$ found in earlier studies was due to the diffusivity of the numerical scheme employed. In any case, the transport is small ($\alpha \sim 10^{-5}$). It is unlikely, therefore, that convection (if present)  provides an appreciable source of disc accretion, and indeed the low $\alpha$, and consequent low dissipation, may explain the inability of convection to sustain itself from its own waste heat.

Recently a further question has arisen: if it is unlikely that convection alone can drive realistic accretion rates, can it perhaps \textit{enhance} accretion in combination with another source of turbulence, such as the MRI? This has proven to be more promising. \cite{hirose2014} and \cite{coleman2018} carried out 3D radiation MHD shearing box simulations with the aim of modelling dwarf novae, and claimed that convection on the left corner of the S-curve's upper branch could enhance $\alpha$ by as much as an order of magnitude, findings which were confirmed independently by \cite{scepi2018a} .\footnote{Simulations with a simple constant thermal diffusivity found similar results \citep{bodo2012}, but were compromised by the severe mismatch between the numerical domain and the disc scale height at saturation (see also the discussion of boundary conditions in \cite{gressel2013}).} This enhancement took place amidst interesting dynamical regimes such as `convective/radiative cycles', chaotic and bursty heat advection, and persistent vertical advection.

These dynamics bring us to the more general question of how convective turbulence and MRI turbulence interact dynamically. On one hand, the disordered motions that comprise MRI turbulence should impede the onset and development of convective eddies (perhaps imposing on them an `effective viscosity'), and on the other hand, MRI motions can transport heat vertically, thus erasing the unstable entropy gradient fueling convection in the first place \citep{gu2000}. For these reasons, identifying convection is itself a challenge, and it is not immediately obvious that it occurs in many of the simulations that claim it does. The potential suppression of convection by MRI motions means that the Schwarzchild criterion is no longer a reliable diagnostic to decide on the presence of convection\footnote{In any case, for the partially ionised conditions relevant for dwarf novae, the Ledoux criterion should really be used.}, and the MRI's vertical transport complicates the attribution of any measured advective heat flux. We expand on these questions in more detail in Section \ref{METHODS_ImportantParametersAndInstabilityCriteria}.
 
\subsection{Motivation and outline}
\label{INTRO_MotivationAndOutline}
Our aim in this paper is to explore the interaction between convection and the MRI and, in particular, to isolate the generic features of their interaction in a controlled manner. We do so through numerical simulations, working in the fully compressible, vertically stratified shearing box approximation. We include both ideal and non-ideal (via an explicit resistivity) magnetohydrodynamics, and employ a perfect gas equation of state and optically thin cooling. We omit much of the complicated physics that has been included in recent work on the subject such as radiative transfer,  a general equation of state, and opacity transitions, thus ensuring that our results are as general as possible. Note that the omission of radiative transfer from our simulations means that the heat flux is always dominated by advection, however.

The structure of the paper is as follows: first, in Section \ref{METHODS} we introduce the basic equations and provide a brief overview of the numerical code, the numerical parameters and set-up, and our main diagnostics. We discuss the criteria for convective instability in Section \ref{METHODS_ImportantParametersAndInstabilityCriteria}. In Section \ref{HEIGHTDEPCOOLING} we investigate the interplay between convection and the MRI in simulations with a height-dependent cooling prescription. We discuss our results in Section \ref{DISCUSSION} and present our conclusions in Section \ref{CONCLUSIONS}. Finally, in the appendices we bring to the fore some subtle but important issues affecting the vertical structure of the disc in vertically stratified shearing box simulations when alternative cooling prescriptions are used (implicit cooling by advection of fluid across the vertical boundaries, and explicit cooling that is uniform in space).

\section{Methods}
\label{METHODS}

\subsection{Governing equations}
\label{METHODS_GoverningEquations}
We work in the shearing box approximation
\citep{goldreich1965,hawley1995,latter2017local},
which treats
a local region of a disc as a Cartesian box located at some fiducial
radius $r = r_0$ and orbiting with the angular frequency of the disc
at that radius $\Omega_0 \equiv \Omega(r_0)$. A point in the box has
Cartesian coordinates $(x, y, z)$ which point in the radial, azimuthal, and vertical directions.
The equations of magnetohydrodynamics are now
\begin{align}
&\partial_t \rho + \nabla \cdot (\rho \mathbf{u}) = 0, \label{SB1} \\
&\partial_t \mathbf{u} + \mathbf{u}\cdot\nabla \mathbf{u} = -\frac{1}{\rho} \nabla P - 2\Omega_0 \mathbf{e}_z \times \mathbf{u} + \mathbf{g}_{\text{eff}} + \frac{1}{\mu_0 \rho}(\nabla\times\mathbf{B})\times\mathbf{B},
\label{SB2} \\
& \partial_t (\rho e) +  \mathbf{u}\cdot \nabla (\rho e) = -\gamma \rho e \nabla\cdot\mathbf{u} + \frac{\eta}{\mu_0}|\nabla\times\mathbf{B}|^2+\Lambda_\text{c} ,
\label{SB3} \\
& \partial_t \mathbf{B} = \nabla\times(\mathbf{u}\times\mathbf{B})+\eta\nabla^2\mathbf{B}, \label{SB4}
\end{align}
with the symbols taking their usual meanings. The system is closed with the caloric equation of state for a perfect gas $P = e
(\gamma - 1) \rho$ where $e$ is the specific internal energy. The adiabatic index (ratio of specific
heats) is denoted by $\gamma$ and is taken to be $5/3$. 
The temperature $T$, when needed, is obtained via the thermal equation of state for a perfect gas $P = (\mathcal{R}/\mu)\rho T$. Here
 $\mathcal{R}$ is the gas constant, and $\mu = 0.5$
is the mean molecular weight. The effective gravitational potential is embodied
in the tidal acceleration
$\mathbf{g}_{\text{eff}} = 2q\Omega_0^2x\hat{\mathbf{x}} -
\Omega_0^2z\hat{\mathbf{z}}$, where $q$ is the dimensionless shear parameter $q \equiv \left.d\ln{\Omega}/d\ln{r}\right\vert_{r=r_0}$. For Keplerian discs $q=3/2$ a value we adopt throughout. In some of our simulations we also employ an explicit resistivity $\eta$. The permeability of free space is denoted by $\mu_0$.

To account for thermal energy loss via radiation we implement a linear cooling term (also known as \textit{beta-cooling}) $\Lambda_\text{c} \equiv - \rho e /\tau_\text{c}$. This lowers the thermal energy in affected cells on a constant cooling time-scale given by $\tau_\text{c}$. In fully turbulent MRI simulations this cooling, together with cooling due to advection of material across the vertical boundaries, is balanced by heating due to the dissipation of MRI turbulence (and resistive or Joule-heating, when explicit resistivity is included). In most of our simulations the beta cooling is $z$-dependent, so as to mock up the effects of an optically thick mid-plane surrounded by an optically thin corona. This is facilitated by turning on $\Lambda_c$ for $|z| > 0.75\,H_0$. Within $|z| < 0.75\,H_0$ the only way for heat to be removed is by vertical advection of fluid. 

\subsection{Important parameters and instability criteria}
\label{METHODS_ImportantParametersAndInstabilityCriteria}

\subsubsection{Criteria for convective instability in an inviscid fluid}
The onset of thermal convection perpendicular to the plane of the disc is controlled by the vertical gradient of specific entropy. The criterion for convective instability (called the \textit{Schwarzschild criterion}) is given by $\partial s/\partial z < 0,$
where $s$ is the specific entropy. The Schwarzschild criterion is most easily and intuitively derived via a parcel argument, see, for example, \cite{landau1987} and assumes only that perturbed fluid elements behave adiabatically (i.e. each fluid element conserves its local entropy). It is valid for a laminar, inviscid fluid of uniform composition (i.e. constant $\mu$) obeying a \textit{general} equation of state, but in the absence of magnetic tension. When the composition of the fluid is non-uniform (as in dwarf novae in outburst, where there are compositional gradients due to the ionisation of hydrogen), the Schwarzschild criterion is generalized to the \textit{Ledoux criterion} (see, for example, \cite{kippenhahn1990}).

In a \textit{perfect gas} the entropy can be related to the pressure and density using $s = c_v \text{ln}(P/\rho^{\gamma})$, where $c_v$ is the specific heat capacity at constant volume. Together with the thermal equation of state for a perfect gas and $g_z = -\Omega_0^2z$, the Schwarzschild criterion can be expressed in terms of pressure and density as
\begin{equation}
N^2_B = z \left[\frac{1}{\gamma} \frac{\partial \ln{P}}{\partial z} - \frac{\partial \ln{\rho}}{\partial z}\right]\Omega_0^2 < 0,
\label{EQUN_BuoyancyFrequencyPerfectGas}
\end{equation}
where $N_B$ is the buoyancy frequency. 
It is important to note that the expression above for $N_B$ is \textit{not} valid for a general equation of state. It is only valid for a \textit{perfect gas} for which $\mu$, $c_p$, $c_v$, and $\gamma$ are assumed to be constants, not always the case in dwarf novae.

\subsubsection{Convective instability in a viscous (or turbulent) fluid}
\label{METHODS_CriteriaForConvectionInAViscidFluid}
In a non-ideal fluid, convection is mitigated by the effects of viscosity $\nu$\ and thermal
diffusivity $\chi$, with the ratio of destabilizing  and stabilizing processes quantified by the \textit{Rayleigh number}
\begin{equation}
\text{Ra} \equiv \frac{\bigl\lvert N_{B}^2 \bigr\lvert H^4}{\nu \chi},
\label{EQUN_RayleighNumber}
\end{equation}
where $H$ is the disc's scale height. Thus in a viscid fluid, the criterion that the square of the buoyancy frequency is negative, i.e. $N_B^2 < 0$, is a necessary condition but not a sufficient one. When explicit viscosity and thermal diffusivity are included, convective instability requires both that $N_B^2 < 0$, \textit{and} that the Rayleigh number exceed some critical value $\text{Ra}_c$.  

The criterion $\text{Ra}> \text{Ra}_c$ applies to a laminar equilibrium with $\nu$ and $\chi$ the \textit{molecular} diffusivities. However, it is tempting to generalise the concept to a \textit{turbulent} background state, such as might be generated by the MRI. Turbulent motions transport both momentum and heat, which may be parametrised by an \textit{effective} (or turbulent) viscosity and thermal diffusivity respectively. The idea that disc turbulence, and the MRI in particular, can transport momentum has been long established, but the ability of turbulence to transfer heat has only been discussed sporadically \citep[e.g.][the last suggesting flows associated with rising flux tubes can transport heat]{rudiger1987,heinrich1994, DAlessio1998,gu2000,blaes2011}. If we combine these effective diffusivities with a fixed $N_B^2$ we may construct an effective or turbulent Rayleigh number $\text{Ra}_\text{eff}$ that quantifies the susceptibility of the turbulent state to convection. It is very likely that the effective Ra is significantly smaller than the laminar/molecular Ra in an MRI unstable disc, and potentially smaller than the relevent $\text{Ra}_c$. Certainly, the sign of $N_B^2$  is insufficient to assign convection to MRI-turbulent flows, as is often done.

We caution, however, that effective diffusion coefficients are just a model for the actual turbulent transport, in particular turbulent fluxes rarely behave behave diffusively on the length and timescales of the principle eddies. A more general, albeit less quantitative, way of thinking is to consider the length and timescales involved. Consider timescales: the turn-over time of the most vigorous MRI eddies will be of the order $\tau_{\text{MRI}} \sim 1 / \Omega$, whereas the fastest growing convective modes grow on timescales of the order $\tau_\text{conv} \sim 1 / N_B > 1/\Omega$ (see Figure 3 in \cite{heldlatter2018}). In addition to sharing similar timescales, the MRI and convective eddies share similar length scales; in fact the radial length scales favoured by convection can be arbitrarily small \citep[see][]{ruden1988axisymmetric,heldlatter2018}, and thus well within the range of the MRI's turbulent cascade. Though these scale comparisons highlight the crudeness of the effective Rayleigh number approach, they nevertheless strengthen the idea that the two phenomena interact significantly, and indeed argue that convective motions, being slower, will become disrupted before they can grow and form coherent plumes. Note that some of these ideas were foreshadowed in the rarely cited \cite{gu2000}.

These concepts can be extended to the nonlinear regime. If convection can get started, despite the difficulties posed by the turbulent background, one might be tempted to treat convection as occurring `on top of' the MRI turbulence. Indeed several authors who have investigated the interplay between the MRI and convection analytically and numerically imply that the two instabilities simply interact in an `additive' manner --- in other words convection and the MRI are operating together simultaneously and independently \citep[e.g.][]{narayan2000self,quataert2000convection,abramowicz2002radial,bodo2012magnetorotational,hirose2014}. But, as we will demonstrate in controlled numerical experiments set-up to sustain both convection and the MRI, the two instabilities \textit{do not} operate in this way: they interact in non-trivial ways (see Section \ref{HEIGHTDEPCOOLING}). 

Lastly, magnetic fields will also impact on the onset and development of convection (e.g. \cite{weiss2014} and references therein). The magnetic tension associated with a large-scale magnetic field impedes nascent convective motions in a relatively straightforward way, but the small-scale magnetic fluctuations associated with the zero-net flux MRI may have an effect that can be packaged away in the effective viscosity discussed above. Through the rest of the paper, we will adopt this interpretation.
 
\subsection{Numerical set-up}
\label{METHODS_NumericalSetUp}

\subsubsection{Code}
\label{Methods_Codes}
For our simulations we use the conservative, finite-volume code \textsc{PLUTO} \citep{mignone2007}. Unless stated otherwise, we employ the HLLD Riemann solver, 2nd-order-in-space linear interpolation, and the 2nd-order-in-time Runge-Kutta algorithm. In addition, in order to enforce the condition that $\nabla\cdot\mathbf{B}=0$ we employ Constrained Transport (CT). Unless stated otherwise we employ the UCT-Contact algorithm to calculate the EMF at cell edges.\footnote{Although we have omitted the details here for brevity, we have found in both isothermal and non-isothermal runs that use of the UCT-HLL algorithm for calculating the EMF at cell edges can mitigate or even kill the MRI entirely.} To allow for longer time-steps, we take advantage of the \textsc{FARGO} scheme \citep{mignone2012}. When explicit resistivity $\eta$ is
included, we further reduce the computational time via the Super-Time-Stepping (STS) scheme \citep{alexiades1996super}. Ghost zones are used to implement the boundary conditions.

We use the built-in shearing box module in \textsc{PLUTO}
\citep{mignone2012}. Rather than solving Equations \eqref{SB1}-\eqref{SB3} (primitive
form), \textsc{PLUTO} solves the governing equations in conservative form, evolving the total energy equation rather than the thermal energy equation. Note that due to the conservative implementation of the equations, kinetic and magnetic energy is not lost to the grid but converted directly into thermal energy.  
 In \textsc{PLUTO} the beta cooling term $\Lambda_\text{c} = -\rho e /\tau_\text{c}$ is not implemented directly in the total energy equation. Instead, it is included on the right-hand-side of the thermal energy equation, which is then integrated (in time) analytically.

\subsubsection{Initial conditions and units}
\label{METHODS_InitialConditionsAndUnits}
All our simulations are initialized from an equilibrium exhibiting a Gaussian
density profile 
\begin{equation}
 \rho = \rho_0 \text{exp}\left[-z^2/(2H_0^2)\right],
 \label{densityinitialization}
\end{equation}
where $\rho_0$ is mid-plane density, and $H_0$ is the scaleheight at the mid-plane at
initialization (formally defined below). 

The background velocity is given by $\mathbf{u} = -(3/2) \Omega_0 x \,
\mathbf{e}_y$. At initialization we usually perturb all the
velocity components with random noise exhibiting a flat power
spectrum. The perturbations $\delta \mathbf{u}$ have maximum
amplitude of about $5\times10^{-2}\,c_{s0}$. Here $c_{s0}$ is the sound speed at the mid-plane at initialization.

All simulations are initialized with a \textit{zero-net-flux} (ZNF) magnetic field configuration: at initialization $\mathbf{B}_0 = B_0\sin{(k_x x)}\mathbf{\hat{e}}_z$ with radial wavenumber $k_x = 2\pi/L_x$. The field strength $B_0$ is controlled through the ratio of gas pressure to magnetic pressure at the mid-plane $\beta_0 \equiv P / (B_0^2 /2)$. In our simulations we set $\beta_0 \equiv 1000$. 

Time units are selected so that $\Omega_0 = 1$. From now the subscript
on the angular frequency is dropped. The length unit
is chosen so that the initial mid-plane sound speed $c_{s0} = 1$, which in turn defines a reference scale-height $H_0\equiv c_{s0} / \Omega_0=1$. 
Note, however, that the sound speed (and the scale-height) is generally a function of both space and time. Finally the mass unit is set by the initial mid-plane density which is $\rho_0 = 1$.

\subsubsection{Box size and resolution}
\label{METHODS_BoxSizeAndResolution}
We measure box size in units of initial mid-plane scale-height $H_0$, defined above. For our fiducial simulations we
employ a resolution of $128\times128\times196$ in boxes of size $4H_0\times4H_0\times6H_0$, which corresponds to a resolution of 32 grid cells per $H_0$ in all directions. 

In order to test the convergence of our results with resolution, in a few select simulations we employ a higher resolution of $256\times256\times392$, corresponding $64$ grid-cells per scale-height in each direction. We also investigate the effect on our results of narrower boxes of radial extent $L_x = H_0$ and $L_x = 2H_0$ (keeping the number of grid-cells per scale-height fixed at $32/H_0$). Selected simulations have also been rerun in taller boxes of vertical extent $L_z = 8H_0$.

\subsubsection{Boundary conditions and mass source term}
\label{METHODS_BoundaryConditions}
We use shear-periodic boundary conditions in the $x$-direction
\cite[see][]{hawley1995} and periodic boundary conditions in the
$y$-direction. In the vertical direction, we keep the ghost zones
associated with the thermal variables in
isothermal hydrostatic equilibrium, in the manner described in
\cite{zingale2002mapping} (the temperature of the vertical ghost zones
is updated at each time-step to match the temperature in the active cells bordering the ghost cells).
For the velocity components we use mostly
standard \textit{outflow} boundary conditions in the vertical
direction, whereby the vertical gradients of all velocity components
are zero and variables in the ghost zones are set equal to those in
the active cells bordering the ghost zones.  For the magnetic field we employ
`vertical field' boundary conditions (i.e. $B_x = 0, B_y = 0, \partial{B_z}/{\partial z} = 0$) \cite[see, for e.g.][]{riols2018}.

To prevent mass loss through the vertical boundaries from depleting the box, we employ a simple mass source term. At the end of the $n$th step, we subtract the total mass in the box at the end of that step $M_n$ from the total mass in the box at initialization $M_0$. This mass difference $\Delta M_n \equiv M_0 - M_n$ is added back into the box with the same profile used to initialize the density (cf. Equation \ref{densityinitialization}). 

We have encountered numerical difficulties at very short cooling timescales when strong magnetic fields are advected across the vertical boundaries \citep[see][]{stonehawley1996}. Thus in some of our simulations (see Section \ref{HEIGHTDEPCOOLING_ParameterSurvey}) we added a thin highly resistive layer (spanning the last three vertical cells in the active domain), wherein the magnetic field is dissipated, in order improve code stability.

\subsection{Diagnostics}
\label{METHODS_Diagnostics}

\subsubsection{Averaged quantities}
\label{METHODS_AveragedQuantities}
The volume-average of a quantity $X$ is denoted $\langle X \rangle$ and is defined as 
\begin{equation}
\langle X \rangle(t) \equiv \frac{1}{V} \int_V X(x, y, z, t) dV
\end{equation}
where $V$ is the volume of the box.

We are also interested in averaging certain quantities (e.g. the Reynolds stress) over time. The temporal average of a quantity $X$ is denoted $\langle{X}\rangle_t$ and is defined as
\begin{equation}
\langle X \rangle_t (x, y, z) \equiv \frac{1}{\Delta t} \int_{t_i}^{t_f} X(x, y, z, t) dt,
\end{equation}
where we integrate from some initial time $t_i$ to some final time $t_f$ and $\Delta t \equiv t_f - t_i$.

If we are interested only in the vertical structure of a quantity $X$ then we average over the $x$- and $y$-directions, only. The horizontal average of that quantity is denoted $\langle X \rangle_{xy}$ and is defined as
\begin{equation}
\langle X \rangle_{xy}(z, t) \equiv \frac{1}{A} \int_A X(x, y, z, t) dA,
\end{equation}
where $A$ is the horizontal area of the box. Horizontal averages over different coordinate directions (e.g. over the $y$- and $z$-directions) are defined in a similar manner.

\subsubsection{Vertical profiles}
\label{METHODS_VerticalProfiles}
We track the vertical profiles of horizontally averaged pressure, density,
temperature, Reynolds stress, magnetic stress, and plasma beta. Of special interest is the buoyancy
frequency, which is calculated from the pressure and density data by finite differencing the formula
\begin{equation}
\left\langle \frac{N_B^2}{\Omega^2} \right \rangle_{xy} = \left \langle z\left [\frac{1}{\gamma}\frac{d\ln{P}}{dz} - \frac{d\ln{\rho}}{dz}\right] \right\rangle_{xy}.
\label{EQUN_BuoyancyFrequencyProfile1}
\end{equation}
Note we calculate horizontal (and time) averages after calculating $N_B^2$. To check the effects of the order of operations we have also calculated the buoyancy frequency by first calculating averages of $P$ and $\rho$ and then calculating $N_B^2$. The percentage difference between the two procedures is less than $1 \%$ within $z = \pm 2H_0$.

We also calculate the (horizontally averaged) vertical profiles of mass and heat flux. We define the mass flux and heat flux\footnote{Note that these expressions are in terms of dimensionless (code) variables. By balancing the caloric EOS $P=(\gamma - 1)\rho e$ with the thermal EOS $P = (\mathcal{R}/\mu)\rho T$ we have obtained an expression for the heat flux $F_z \equiv \rho e u_z$ in terms of the temperature, i.e. $F_z \propto \rho u_z T$. The constant of proportionality has been absorbed into the definition of the dimensionless heat flux.}
\begin{equation}
F_{\text{mass}} = \langle \rho  u_z \rangle_{xy}, \qquad F_{\text{heat}} = \langle \rho u_z  T \rangle_{xy}.
\end{equation}

\subsubsection{Reynolds and magnetic stresses, and alpha}
\label{METHODS_ReynoldsAndMagneticStressesAndAlpha}
In accretion discs, the radial transport of angular momentum is
related to the $xy$-component of the total stress
\begin{equation}
\Pi_{xy} \equiv R_{xy} + M_{xy},
\label{totalstress}
\end{equation}
in which $R_{xy} \equiv \rho u_x \delta u_y$ is the Reynolds stress, where $\delta u_y \equiv u_y + q\Omega x$ is the fluctuating part of the y-component of the total velocity $u_y$, and $M_{xy} \equiv -B_x B_y$ is the magnetic stress. The total stress is related to the classic dimensionless
parameter $\alpha$ by
\begin{equation}
\alpha \equiv \frac{\langle \Pi_{xy} \rangle}{\langle P \rangle},
\label{alpha}
\end{equation}
Note that some definitions of alpha include the dimensionless shear parameter in the denominator above. To compare our results more easily with the literature, we drop this factor.

\subsubsection{Energetics}
\label{METHODS_EnergyDensities}
The total energy density is given by
\begin{equation}
E_\text{total} = \frac{1}{2}\rho u^2 + \frac{1}{2}B^2 + \rho \Phi + \rho e,
\end{equation}
where the terms on the right-hand side correspond to the kinetic $E_{\text{kin}}$, magnetic $E_{\text{mag}}$, gravitational potential, and thermal $E_{\text{th}}$ energy densities, respectively. Here $\Phi = \frac{1}{2}\Omega_0^2z^2-\frac{3}{2}\Omega_0^2x^2$ is the effective gravitational potential in the shearing box approximation for a Keplerian disc. 

Because we employ open boundaries in the vertical direction, energy is lost through advection of fluid across the vertical boundaries (which we refer to as \textit{box-cooling}). It is important to determine the extent to which this energy loss influences the total energy budget. The integrated flux of total energy across the vertical boundaries is given by
\begin{equation}
\mathcal{F}_z \equiv \frac{1}{V}\left[\iint (u_z E_\text{tot} + u_z P_t - B_z(\mathbf{u}\cdot\mathbf{B}))\,dxdy\right]_{z = -L_z/2}^{z = +L_z/2},
\end{equation}
where $P_t = P + B^2/2$.

The integrated flux of total energy across the vertical boundaries (normalized by the volume-averaged thermal energy density) can be used to estimate the rate at which energy is lost through the vertical boundaries, the inverse of which is known as the wind cooling time $\tau_\text{w}$ \citep[see][]{riols2018}, and this is defined by
\begin{equation}
\frac{1}{\tau_\text{w}(t)} \equiv \frac{\mathcal{F}_z}{\langle E_\text{th}\rangle}.
\label{EQUN_WindCoolingRate}
\end{equation}

\subsubsection{2D Power Spectra}
\label{METHODS_2DPowerSpectra}
In order to distinguish convection from the MRI in a quantitative manner we use the 2D power spectrum of the specific vertical kinetic energy. To calculate this, we first extract the ($y$-averaged) vertical component of the velocity $w \equiv \langle u_z \rangle_y$ between $z=-H_0$ and $z=+H_0$ (as most of the activity in the disc lies within this region).

Because we use open boundary conditions (in $z$) and shear periodic boundary conditions (in $x$) this data is \textit{not} periodic in either direction. To make it periodic we reflect the data in $x$ and $z$ to create a doubly-periodic array. The $kl$th component of the 2D discrete Fourier transform of this data is then defined as
\begin{equation}
\widehat{w}_{kl} = \sum_{n = 0}^{\tilde{N}_x-1} \sum_{m = 0}^{\tilde{N}_z-1} w_{mn}\,\text{exp}\left\{-2\pi i \left(\frac{mk}{\tilde{N}_x}+\frac{nl}{\tilde{N}_z} \right)\right\},
\end{equation}
where $k=0,\ldots,\tilde{N}_x-1$, $l=0,\ldots,\tilde{N}_z-1$, and $w_{mn}$ denotes the $mn$th component of the 2D array $w \equiv \langle u_z \rangle_y$. Note that  $\tilde{N}_x$  (and $\tilde{N}_z$) are the number of radial (vertical) cells in the extended periodic partition of the disc; in general they are \textit{not} equal to the total number of radial (vertical) cells in the domain $N_x$ (and $N_z$).

The power in the specific vertical kinetic energy can then be obtained from
\begin{equation}
\hat{E}_{\text{kin,z}} = \frac{1}{2} \bigl\lvert \widehat{w} \bigr\rvert^2,
\label{EQUN_VerticalKEPower}
\end{equation}
where $ \bigl\lvert \widehat{w} \bigr\rvert^2 \equiv \widehat{w} \widehat{w}^{*}$. Finally we plot Equation \ref{EQUN_VerticalKEPower} in the $(k_x, k_z)$-plane to obtain the 2D power spectrum of the specific vertical kinetic energy.

\subsubsection{Detecting convection}
\label{METHODS_DiagnosticsForConvection}
In general it is difficult to detect convection against the backdrop of MRI turbulence. To aid us in determining whether convection is present in our simulations or not, we employ various diagnostics. Our first diagnostic is the sign of the buoyancy frequency $N_B^2$ (Equation \ref{EQUN_BuoyancyFrequencyProfile1}), though as discussed in Section \ref{METHODS_CriteriaForConvectionInAViscidFluid} this is a necessary but not a sufficient criterion in turbulent flows. Thus where $N_B^2 > 0$ we can definitely rule out the presence of convection.  Another common diagnostic is the vertical heat flux, or some measure thereof.

Another approach is visual inspection of the flow field in the $xz$-plane (i.e. that the vertical velocity $u_z$ exhibit hot updrafts and cool downdrafts). To try to quantify any vertical `structure' that we pick up visually, we use a third diagnostic, namely  the 2D power spectra outlined in Section \ref{METHODS_2DPowerSpectra}. From linear theory, convective cells are arranged such as to minimize radial fluid motion (i.e. $k_x/k_z \gg1$), whereas the MRI tends to \textit{maximize} radial fluid motions (i.e. $k_x/k_z \ll 1$). Thus each instability conveniently inhabits different regions of wavenumber space. 

A fourth diagnostic is the time-evolution of the ratio of the vertical to radial kinetic energy density (i.e. $E_{\text{kin},z} / E_{\text{kin},x}$). As we show later, the MRI itself is rather adept at moving fluid vertically thus the vertical kinetic energy density on its own is not a very useful diagnostic for distinguishing convection from the MRI. A more useful diagnostic, however, is the relative amount of vertical motion compared to radial motion. We have found that when convection is dominant, the ratio of vertical to radial kinetic energy is significantly larger than in normal MRI.

Finally, we have found in our hydrodynamic simulations that convection exhibits a small mass flux \textit{towards} the mid-plane \citep[see Figure 8 of][]{heldlatter2018}. Thus convection rearranges the vertical disc structure (by changing the temperature and density profiles) so that the disc is closer to a state of marginal stability. The mass flux, however, is a slow transient, and appears to decrease over very long timescales. Nevertheless over the timescales of interest (c. $100$ orbits) we have found it to be a reliable diagnostic for convection.\footnote{The unforced hydro simulations in question did not include a mass source term, thus we are confident that the inward mass flux observed when convection is present is not an artefact of our mass source term.}

\section{Simulations with height-dependent explicit cooling}
\label{HEIGHTDEPCOOLING}

\subsection{Motivation}
\label{HEIGHTDEPCOOLING_Motivation}

In this section we explore the interplay between the MRI and convection. We include a height-dependent explicit cooling using the linear cooling prescription described in Section \ref{METHODS_GoverningEquations}). The setup by construction gives an advective region around the mid-plane surrounded by a `radiative' region, thus imitating the vertical structure of dwarf novae discs in the high-state, in which an optically thick disc (in which accretion is driven by MRI turbulence, and, possibly, enhanced by convection) is surrounded by an optically thin atmosphere. 

Height-dependent cooling is not the simplest possible set-up, and we have also run vertically stratified non-isothermal simulations that (i) cool only by advection of material across the vertical boundaries (which we term `box-cooling': see Appendix \ref{BOXCOOLING}), and (ii) that employ uniform explicit cooling (Appendix \ref{UNIFORMCOOLING}). However, we find that these simulations lead to a convectively stable entropy profile. More worrying, we find that these set-ups suffer from various pathologies that are alleviated when height-dependent cooling is employed (provided that the explicit cooling timescale is less than the wind-cooling timescale (see Equation \ref{EQUN_WindCoolingRate})).

As discussed in Section \ref{METHODS_CriteriaForConvectionInAViscidFluid}, the onset of convection amidst an MRI turbulent background likely depends not just on the sign of the entropy gradient, but also on the effective Rayleigh number $\text{Ra}_{\textrm{eff}} = N_B^2 H^4 / (\nu_{\textrm{eff}} \chi_\textrm{{eff}})$. This quantity provides us with a means to adjust the conditions that favour convection: at a sufficiently large $\text{Ra}_{\textrm{eff}}$ we expect the onset of convection to be more likely. One way to increase $\textrm{Ra}_{\textrm{eff}}$ may be achieved by \textit{decreasing} the effective/turbulent MRI viscosity $\nu_{\textrm{eff}}$ and thermal diffusivity $\chi_\textrm{{eff}}$. These may be reduced by simply weakening the MRI activity, which we do by employing a small but finite \textit{explicit resistivity} $\eta$. The MRI becomes sluggish the greater the $\eta$; consequently the MRI transports less momentum and heat, the effective diffusivities diminish, and the effective Rayleigh number increases. 

A second way to boost the effective Rayleigh number is by increasing the magnitude of $N_B^2$. We achieve this by decreasing the \textit{cooling timescale} $\tau_\textrm{c}$ in the radiative regions. This strengthens the temperature difference between these upper regions and the mid-plane, which in turn intensifies the unstable entropy gradient in that region. Thus by specifying both the cooling timescale $\tau_{\textrm{c}}$ and an explicit resistivity $\eta$ in our simulations, we can effectively control the strength of convection and of the MRI, respectively, allowing us to explore their interaction in different regimes of parameter space.

\subsection{Set-up}
\label{HEIGHTDEPCOOLING_SetUp}
In the simulations described in this section, we employ an explicit resistivity $\eta$ (which is kept constant in space and time in any individual simulation). We also implement a piecewise-in-space cooling, i.e. cooling is turned on only within a region $|z| > 0.75H_0$ above the mid-plane. This allows for an unstable entropy gradient to develop in the vicinity of the mid-plane. Fluid can still leave the box via the vertical boundaries, but we expect cooling to be dominated by the explicit cooling term rather than by box-cooling, alleviating the issues with vertical heat transport that plague box-cooled runs (see Appendix \ref{BOXCOOLING}). The run with the longest cooling timescale and smallest resistivity (simulation NSTRMC44f1 with $\tau_\text{c} = 10$ orbits, $\eta = 10^{-5}$) is closest to the box cooled run, but the mass loss per step is an order of magnitude smaller and the wind cooling timescale is $\tau_\text{w} \sim 258\,\Omega^{-1}$ compared to the explicit cooling timescale of $\tau_\text{c} \sim 63\,\Omega^{-1}$. Box cooling only becomes more subdominant as the cooling timescale is lowered: at a cooling timescale of $\tau_\text{c} = 1$ orbits (simulation NSTRMC44c6) the mass loss is two orders of magnitude smaller than in the box cooled run, and the wind cooling timescale is $\tau_\text{w} \sim 1693\,\Omega^{-1}$ compared to the explicit cooling timescale of $\tau_\text{c} \sim 6 \,\Omega^{-1}$. Thus box cooling plays no role in the determining the vertical structure of the disc in these simulations.

We initialize all simulations from around orbit $5$ of the non-linear MRI turbulent state of our fiducial (box-cooled) simulation (see Section \ref{BOXCOOLING}). This is sufficiently close to non-linear saturation of the linear MRI that the disc has not had time to heat up sufficiently to fill the box. At initialization the mid-plane scale height is $H\sim1.55H_0$. Each simulation was run for 100-200 orbits.

\begin{figure}
\centering
\includegraphics[scale=0.24]{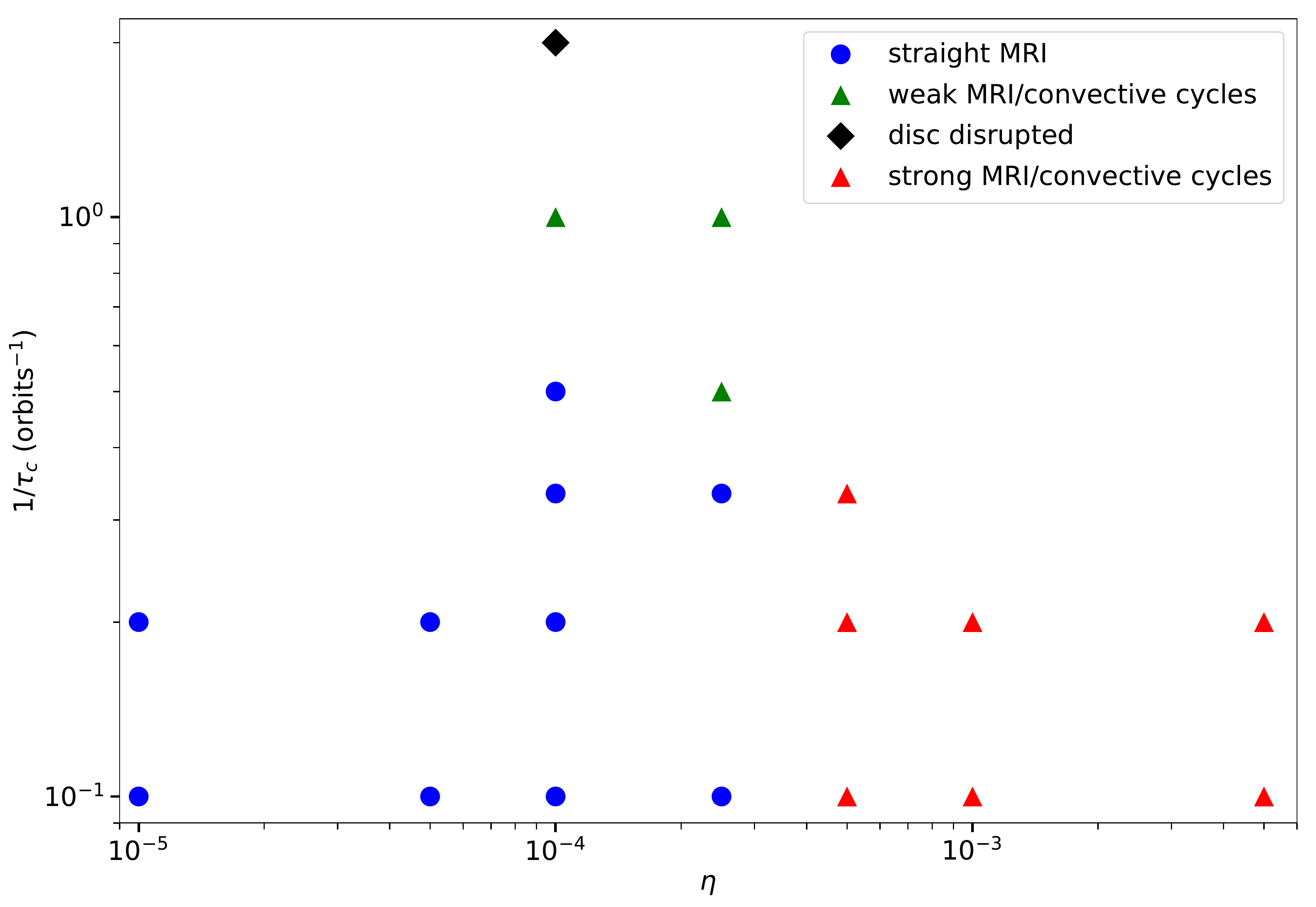}
\caption{Parameter space of simulations run with beta cooling (plotted on the ordinate as the inverse of the cooling timescale $\tau_\textrm{c}$) and explicit resistivity $\eta$. Blue dots correspond to simulations that exhibited MRI only (see Section \ref{HEIGHTDEPCOOLING_MRIOnly}), and red triangles correspond to simulations that exhibited MRI/convective cycles (see Sections \ref{HEIGHTDEPCOOLING_StrongCycles}-\ref{HEIGHTDEPCOOLING_WeakCycles}).}
\label{FIGURE_ParameterSurvey}
\end{figure}

\subsection{Parameter survey}
\label{HEIGHTDEPCOOLING_ParameterSurvey}
Motivated by the arguments presented in Section \ref{HEIGHTDEPCOOLING_Motivation}, in this section we present a parameter survey of simulations run with different combinations of cooling timescale $\tau_\text{c}$ and explicit resistivity $\eta$ (equivalently, magnetic Reynolds number Rm $\equiv c_{s0}H_0/\eta$). 

The results of this survey are plotted in Figure \ref{FIGURE_ParameterSurvey} which shows simulations in the space of inverse cooling timescale $1/\tau_{\text{c}}$ against explicit resistivity $\eta$. We find three distinct outcomes: at high resistivities and long cooling timescales we find that the system exhibits alternating phases in which either the MRI or convection dominates the flow. We refer to these solutions as \textit{strong} MRI/convective cycles (red diamonds in Figure \ref{FIGURE_ParameterSurvey}). Remarkably, during the MRI phases $\alpha$ is enhanced by nearly an order of magnitude. Based on our effective Rayleigh number argument, we expect to recover similar behaviour at more moderate resistivities (i.e. larger effective viscosities) but shorter cooling timescales (i.e. larger entropy gradients). This is indeed what we find, but the character of the cycles is sufficiently altered that we classify them separately: we refer to these solutions as \textit{weak} MRI/convective cycles (green diamonds in Figure \ref{FIGURE_ParameterSurvey}). At low resistivities ($\eta \leq 2.5\times10^{-4}$) and long cooling timescales ($\tau_\text{c} \sim 10$ orbits) we observe only the MRI, essentially recovering the results of our box-cooled and uniformly-cooled simulations, though we emphasize that marginal cases ($\eta \sim 2.5\times10^{-4}$ in particular) are difficult to classify, because on the one hand they do not exhibit bursts in the volume-average of $\alpha$ characteristic of MRI/convective cycles, but on the other hand we do sporadically see some vertical structure in the flow and also in the 2D spectra.

In addition to the existence of these cycles, and the large enhancement in $\alpha$ sometimes observed, a key finding is that there is \emph{no} evidence that the MRI and convection coexist (in the sense that the two instabilities behave in an additive manner, with convection `sitting on top of' MRI turbulence) in any of the three regimes. Each of these three cases (strong MRI/convective cycles, weak MRI/convective cycles, and MRI-only) is discussed in greater detail in Sections \ref{HEIGHTDEPCOOLING_StrongCycles}, \ref{HEIGHTDEPCOOLING_WeakCycles}, and \ref{HEIGHTDEPCOOLING_MRIOnly}, respectively. 

\begin{figure}
\centering
\includegraphics[scale=0.23]{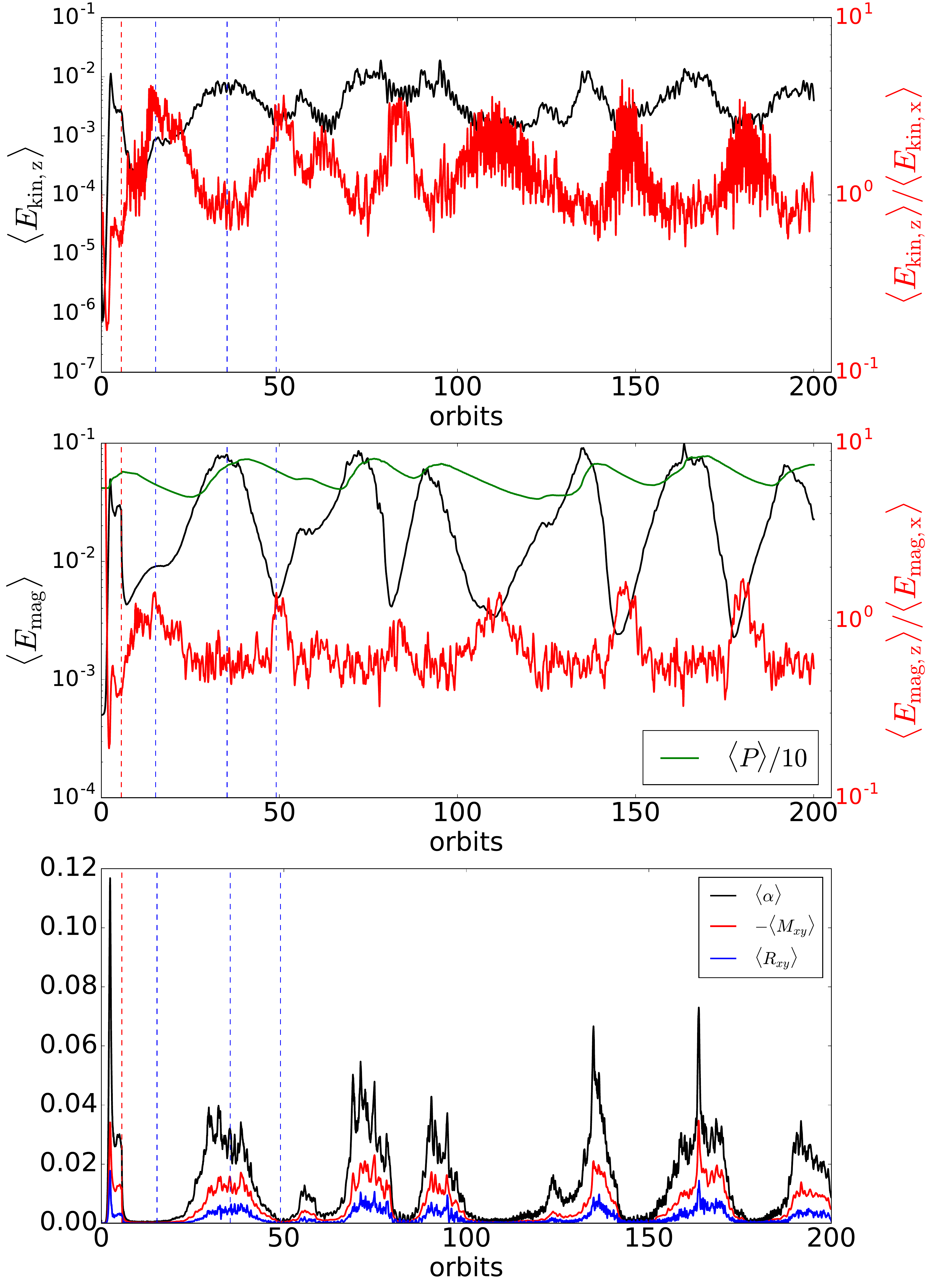}
\caption{Top: semi-log plot of the time-evolution of volume-averaged vertical kinetic energy density (in black) and ratio of vertical to radial kinetic energy (in red) from a simulation exhibiting MRI/convective cycles (NSTRMC44e1). Middle: time-evolution of volume-averaged magnetic energy density (black),  and ratio of vertical to radial magnetic energy (in red). The thermal pressure (divided by 10) is superimposed in green. Bottom: time-evolution of volume-averaged $\alpha$ parameter (black), magnetic stress (red), and Reynolds stress (blue). The dashed vertical red line corresponds to the time at which the resistivity was turned on. Convection-dominated phases are associated with troughs in $\alpha$ while MRI-dominated phases are associated with bursts in $\alpha$ (see Figure \ref{FIGURE_MRISimsWithCoolingFlowFieldSnapshots}). The times at which the snapshots in that figure were taken are indicated by the dashed blue lines in the figure above.}
\label{FIGURE_MRISimsWithCoolingEnergies}
\end{figure}

\subsection{Strong MRI/convective cycles}
\label{HEIGHTDEPCOOLING_StrongCycles}
For a cooling timescale of $\tau_\text{c} = 10$ orbits and resistivities $\eta \geq 5\times10^{-4}$ we observe a state characterized by clear cyclical outbursts in which the flow appears to switch between convection and the MRI. The large enhancement in $\alpha$ during the outbursts, and the clear distinction between convective and MRI-dominated phases, leads us to label these solutions \textit{strong} MRI/convective cycles. As our fiducial example of a run exhibiting this behavior, we have chosen simulation NSTRMC44e1b, which was run with a cooling timescale of $\tau_\text{c}=10$ orbits and an explicit resistivity of $\eta=5\times10^{-3}$. Although the resistivity is relatively large, this simulation exhibits the clearest behavior.

\begin{figure*}
\centering
\subfloat[]{\includegraphics[scale=0.22]{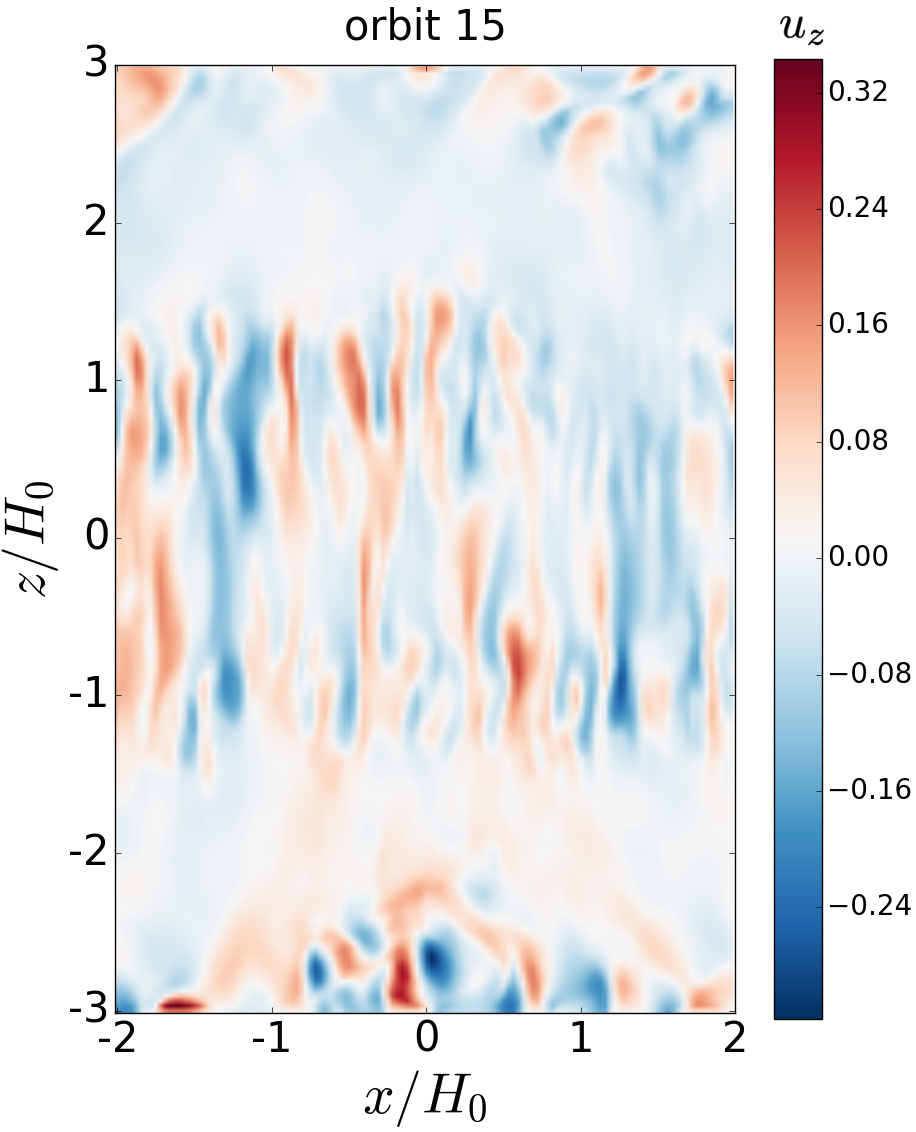}}\hspace{0.04em}
\subfloat[]{\includegraphics[scale=0.22]{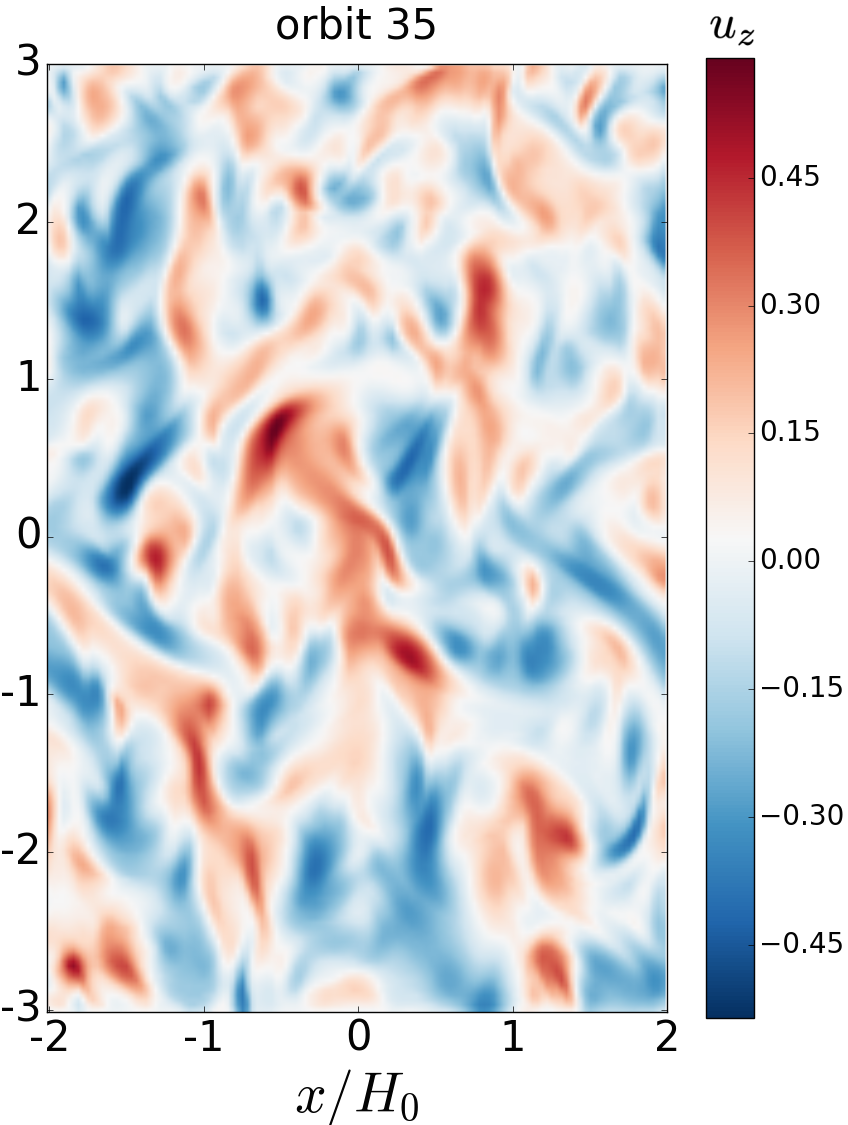}}\hspace{0.04em}
\subfloat[]{\includegraphics[scale=0.22]{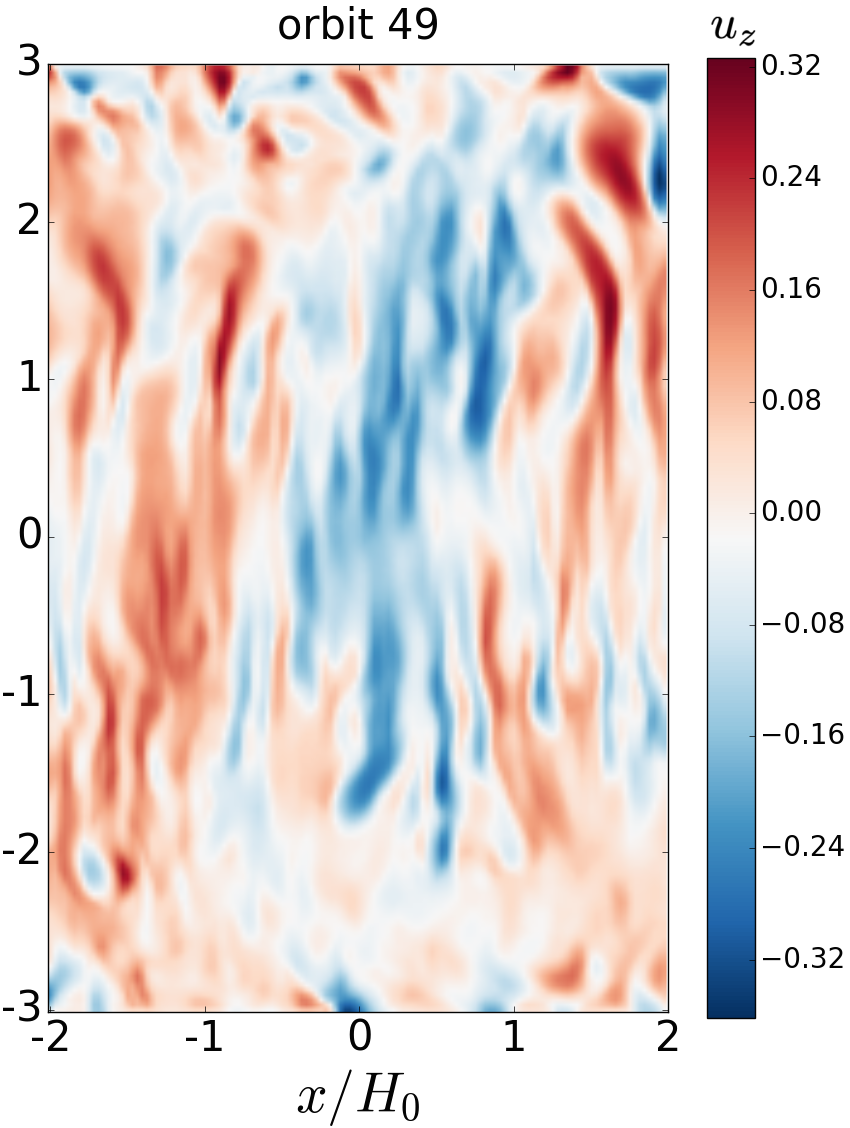}}\hspace{0.04em}
\caption{Snapshots of the vertical component of the velocity in the $xz$-plane taken from snapshots at different times from a simulation exhibiting MRI/convective cycles (NSTRMC44e1). (a) convection-dominated phase, (b) MRI-dominated phase, (c) convection-dominated phase with large-scale convective cells. The simulation employed a constant cooling timescale of $\tau_\text{c}=10$ orbits above $|z| = 0.75H_0$, and a uniform explicit resistivity of  $\eta = 5\times10^{-3}$.}
\label{FIGURE_MRISimsWithCoolingFlowFieldSnapshots}
\end{figure*}

In Figure \ref{FIGURE_MRISimsWithCoolingEnergies} we plot the time evolution of the vertical kinetic energy density (black curve, top panel) and magnetic energy density (black curve, middle panel). Compared to the behavior of the kinetic energy in our box-cooled simulation (see Appendix \ref{BOXCOOLING_TimeEvolutionOfAveragedQuantities}) in which $\langle E_{\text{kin},z} \rangle$ fluctuated around $10^{-3}$ in the non-linear phase, in this simulation we instead observe bursts in the kinetic energy density during which $\langle E_{\text{kin},z} \rangle$ can reach values as high as $10^{-2}$. Between the outbursts, the kinetic energy density is found to be highly oscillatory, reminiscent of the behavior of the kinetic energy observed in simulations of forced compressible (hydro) convection in \cite{heldlatter2018}. To more clearly track the changes in the vertical kinetic energy, we have also plotted the time evolution of the ratio of vertical to radial kinetic energy (red curve, top panel). This is anti-correlated with magnetic and vertical kinetic energies. The vertical kinetic exceeds the radial kinetic energy during convection dominated periods (see below), while during MRI-dominated phases the two are in equipartition. An important point here, however, is that the MRI itself naturally gives rise to stronger vertical flows than convection (as evidenced by the peaks in the vertical kinetic energy during the MRI-dominated bursts in $\alpha$). Finally we also plot the ratio of vertical to radial magnetic energy (red curve, middle panel). The vertical field exceeds the radial field between outbursts (convection-dominated phases), and is less than the radial field during the outbursts (MRI-dominated phases).

In the bottom panel of Figure \ref{FIGURE_MRISimsWithCoolingEnergies} we plot the time evolution of the stresses and of alpha. The initial spike in $\alpha$ around orbit 2 is due to the linear MRI, followed by non-linear saturation. The vertical dashed line indicates the time at which resistivity was turned on. The stress abruptly drops as resistivity appears to quench the non-linear MRI, a fact that we confirm in a simulation in which we turn off the cooling which we describe in Section \ref{DISCUSSION_EffectOfResistivity}. However this quenching is followed by alternating periods of low stress and high stress (`outbursts'). During outburst $\alpha$ can reach as high as $0.08$ (we have observed values up to $0.1$ in some of our preliminary simulations), though typical values of $\alpha$ in outburst are around $\langle \langle \alpha \rangle \rangle \sim 0.04-0.06$.

\subsubsection{Flow structure}
\label{MRIConvectiveCycles_StructureOfTheFlow}
The periods of low stress and of outburst observed in the time-evolution of volume-averaged quantities (see Figure \ref{FIGURE_MRISimsWithCoolingEnergies}) correspond to \textit{convection}-dominated and \textit{MRI}-dominated phases, respectively. This can be seen most clearly in the Figure \ref{FIGURE_MRISimsWithCoolingFlowFieldSnapshots} were we plot the $u_z$ component of the velocity in the $xz$-plane. In the left-most panel, taken from a snapshot during the first \textit{low stress} phase (around orbit 15), the onset of convective instability is clearly visible within about $\pm H_0$ of the mid-plane through thin convective cells consisting of hot rising fluid (in red) and cool sinking fluid (in blue). These thin convective cells are reminiscent of linear convective instability observed in our hydro simulations (see Figure 7 of \cite{heldlatter2018}). Although at this point in the simulation the MRI has been quenched by the explicit resistivity, the combination of cooling from above and residual heating from before the MRI was quenched has triggered what appears to be the linear phase of the convective instability.\footnote{Note that we find that the mid-plane scale-height oscillates in tandem with the stress. Thus it is also possible that the heat source for convection is due to contraction of the disc rather than residual heat from the previous MRI phase.} Conversely, during \textit{outbursts} (middle-panel of Figure \ref{FIGURE_MRISimsWithCoolingFlowFieldSnapshots} taken at orbit 35) the flow is dominated by MRI turbulence. No convective cells are visible in the flow field during outbursts, though the buoyancy frequency is still negative during these phases. Finally, during subsequent low stress (i.e. convection-dominated) phases we see the emergence of large scale convective cells (right-most panel of Figure \ref{FIGURE_MRISimsWithCoolingFlowFieldSnapshots}, taken from a snapshot at orbit 49). These are reminiscent of the large-scale cyclical convective cells reported in hydrodynamic simulations of non-linear forced compressible convection \citep[see Figure 13 of][]{heldlatter2018}. The cells are cyclically created and reformed with the opposite orientation, and their manifestation in the flow field coincides with short term oscillations (with periods of order 1 orbit) in the vertical kinetic energy and in the stresses. However here the heating is supplied self-consistently via dissipation of MRI turbulence from the preceding outburst.

As the cells in the right-hand panel of \ref{FIGURE_MRISimsWithCoolingFlowFieldSnapshots} appear to span the entire vertical extent of the box, we have rerun this simulation in a taller box spanning $\pm 4H_0$ about the mid-plane. We again observe alternating phases of low and high stress, and the flow field exhibits similar behavior during each phase as it does in the simulation run in a shorter box of $\pm 3H_0$ about the mid-plane. In this taller box run, however, the cells can clearly be seen to peter out well before they reach the vertical boundaries.

\begin{figure}
\captionsetup[subfigure]{labelformat=empty}
\centering
\includegraphics[scale=0.44]{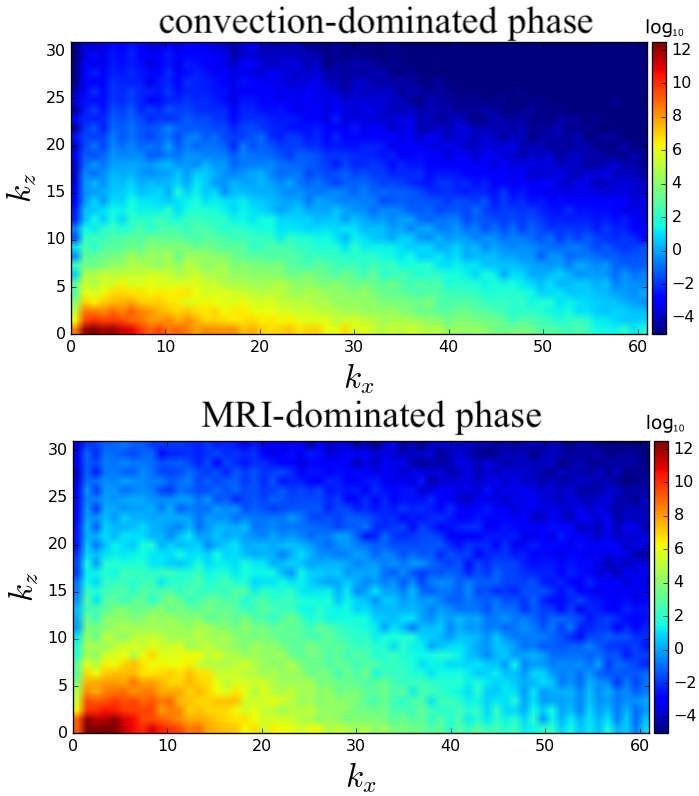}
\caption{2D y-averaged power spectra of specific vertical kinetic energy from a simulation exhibiting MRI/convective cycles (NSTRMC44e1). The colorbars are logarithmic. Top: spectrum taken from snapshots time-averaged over a \textit{convection}-dominated phase. Bottom: same but time-averaged over an MRI-dominated phase. The simulation employed a constant cooling timescale of $\tau_{\text{c}}=10$ orbits above $|z| = 0.75H_0$, and a uniform explicit resistivity of  $\eta = 5\times10^{-3}$. For each spectrum the data has been taken within $\pm H_0$ of the mid-plane and averaged in time over several tens of orbits.}
\label{FIGURE_2DPowerSpectra}
\end{figure}

These visual observations are confirmed by the 2D power spectra of the specific vertical kinetic energy (see Figure \ref{FIGURE_2DPowerSpectra}). During convection-dominated phases (a time-average over one of which is shown in the top panel), the kinetic energy is clearly distributed towards $k_z=0$, representative of the updrafts and downdrafts that dominate the flow structure within around $\pm2H_0$. During MRI-dominated phases on the other hand, the energy is distributed more isotropically in the ($k_x, k_z$) plane.

\begin{figure}
\captionsetup[subfigure]{labelformat=empty}
\centering
\includegraphics[scale=0.16]
{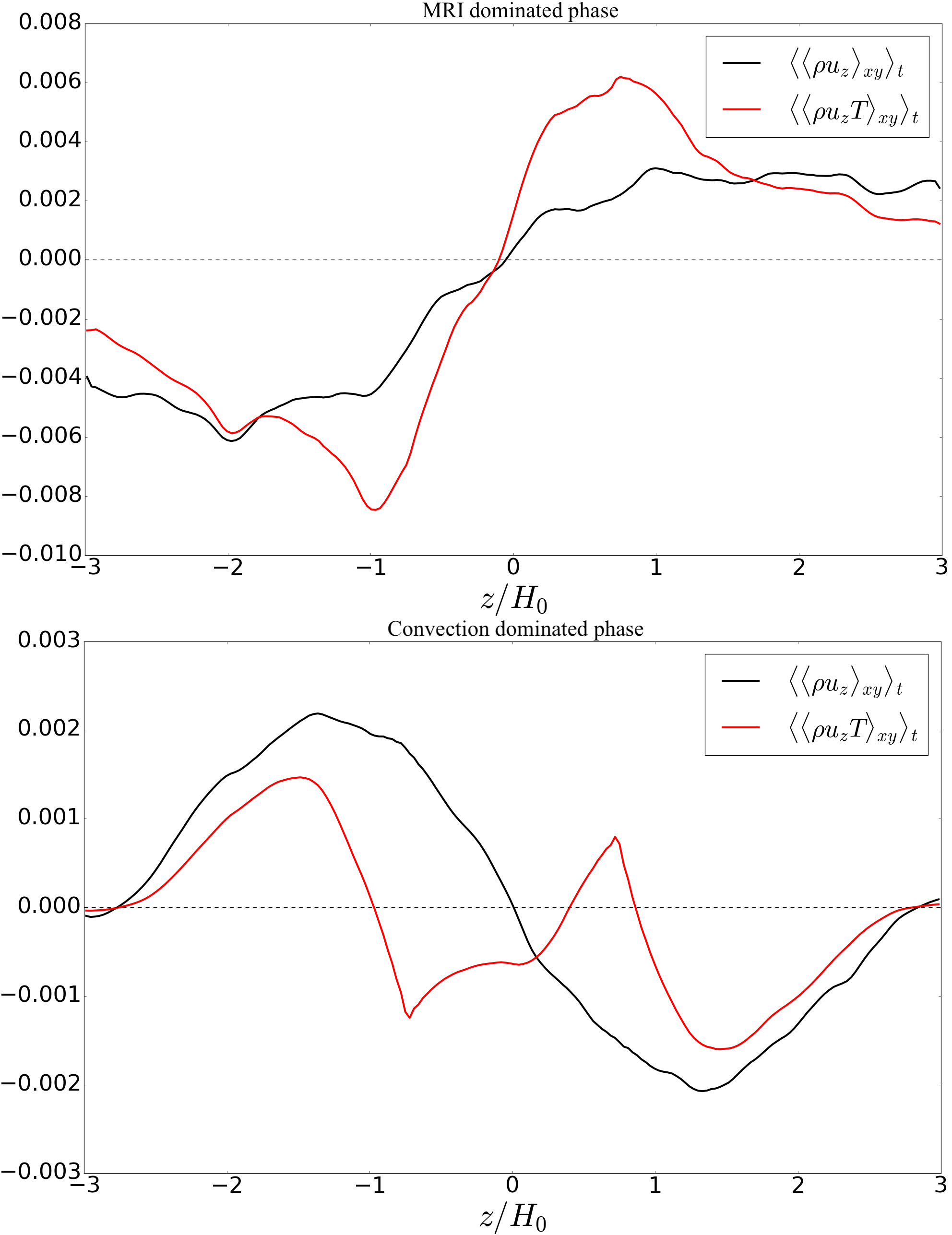}
\caption{Vertical disc structure from a simulation exhibiting MRI/convective cycles (NSTRMC44e1). Top: horizontal- and time-averaged vertical mass and heat flux (top panel) taken from an MRI-dominated phase (orbits 30 to 37). Bottom: the same diagnostics but taken from a \textit{convection}-dominated phase (orbits 144 to 153). The simulation was run with a constant cooling timescale of $\tau_\text{c}=10$ orbits above $|z| = 0.75H_0$, and a uniform explicit resistivity of  $\eta = 5\times10^{-3}$. }
\label{FIGURE_MRISimsWithCoolingVerticalProfiles}
\end{figure}

\subsubsection{Vertical profiles of the disc}
\label{MRIConvectiveCycles_VerticalStructureOfTheDisk}
In Figure \ref{FIGURE_MRISimsWithCoolingVerticalProfiles} we show the vertical profiles (averaged over the $x$- and $y$- directions and over time) of the vertical mass and heat fluxes. The top panel consists of profiles time-averaged over snapshots from an MRI-dominated phase (orbits 30 to 37), whereas the lower panel consists of profiles time-averaged over snapshots from a convection-dominated phase (orbits 144 to 153). There are both qualitative and quantitative differences in the vertical profiles compared to those measured in our box-cooled MRI simulation (see Figure \ref{FIGURE_FiducialSimsVerticalProfiles}).

During the MRI-dominated phase (upper panel of Figure \ref{FIGURE_MRISimsWithCoolingVerticalProfiles}) the heat and mass flux are both directed outwards, consistent with the transport properties of MRI turbulence (see Figure \ref{FIGURE_FiducialSimsVerticalProfiles} taken from our box-cooled simulation). Unlike in our box-cooled simulation, however, the buoyancy frequency (not shown) is negative within $\pm H_0$ of the mid-plane, reaching a minimum value of around $\text{min}(\langle \langle N_B^2\rangle_{xy} \rangle_t) \sim -0.07$ around $|z| \sim 0.75H_0$. As we have emphasized, however, $N_B^2 < 0$ does not automatically imply convection, and thus the negativity of the buoyancy frequency during the MRI-dominated phase does not mean that there is convection occurring in tandem with the MRI.

In the \textit{convection}-dominated phase (lower panel) the vertical heat flux is also directed away from the mid-plane but the vertical mass flux is directed \textit{towards} the mid-plane. We have found this inward mass flux to be a reliable, if transient, sign of the onset of convection, and taken together with the colorplots of the flow field and 2D power spectra, it gives us some confidence that our diagnostics can distinguish between MRI and convective turbulence. (Note that contraction of the disc was also observed during the convective epochs of \cite{hirose2014} and \cite{coleman2018}.) The buoyancy frequency during the convection-dominated phase is negative within $\sim1.2 H_0$, but its minimum value is twice what it is in the MRI-dominated phase with $\text{min}(\langle \langle N_B^2\rangle_{xy} \rangle_t) \sim -0.12$ around $|z| \sim 0.75H_0$. Finally, the temperature profile time-averaged over the duration of the simulation (orbit 50 to orbit 200; not shown), drops monotonically from the mid-plane compared to the flat topped, isothermal profile observed within around $\pm2H_0$ of the mid-plane in our box-cooled simulation. 

An interesting feature is that the vertical heat flux during the convective-dominated phase actually appears to be less than that during the MRI-dominated phase, suggesting that turbulent convection is not so efficient at transporting heat as to completely remove the residual heat left over from the previous MRI-dominated phase. For example during the third MRI burst (orbits 89 to 95) the maximum vertical heat flux is max$(\langle \langle F_z \rangle_{xy} \rangle_t) \sim 0.0033$, whereas during the subsequent convection-dominated phase (orbits 103 to 115) the maximum heat flux is just max$(\langle \langle F_z \rangle_{xy} \rangle_t) \sim 0.0009$. Conversely, the buoyancy frequency is consistently higher when averaged over the convection-dominated phases than when averaged over the MRI-dominated phases: min$(\langle \langle N_B^2\rangle_{xy} \rangle_t) \sim -0.13$ during the convection-dominated phase (orbits 103 to 115), compared to min$(\langle \langle N_B^2\rangle_{xy} \rangle_t) \sim-0.04$ during the preceding MRI dominated phase (orbits 89 to 95). This pattern, of large (minimum, i.e. most negative) $N_B^2$ but small heat flux when convection is dominant and small (minimum) $N_B^2$ but large heat flux when the MRI is dominant, is one that we find consistently in our simulations, including those run at shorter cooling timescales (see Section \ref{HEIGHTDEPCOOLING_WeakCycles} and Table \ref{TABLE_MRISimulationsWithCooling}). We caution, however, that taking meaningful time-averages is difficult here, first due to the relatively short-intervals over which the time-averages have been taken (because the phases themselves only last several tens of orbits at best), and second because the disc is not in thermal equilibrium in either phase. During the MRI bursts heat is being built up due to the dissipation of MRI turbulence, whereas during the convective-dominated phase heat is no longer being built up but is being redistributed.

\begin{figure}
\centering
\includegraphics[scale=0.21]{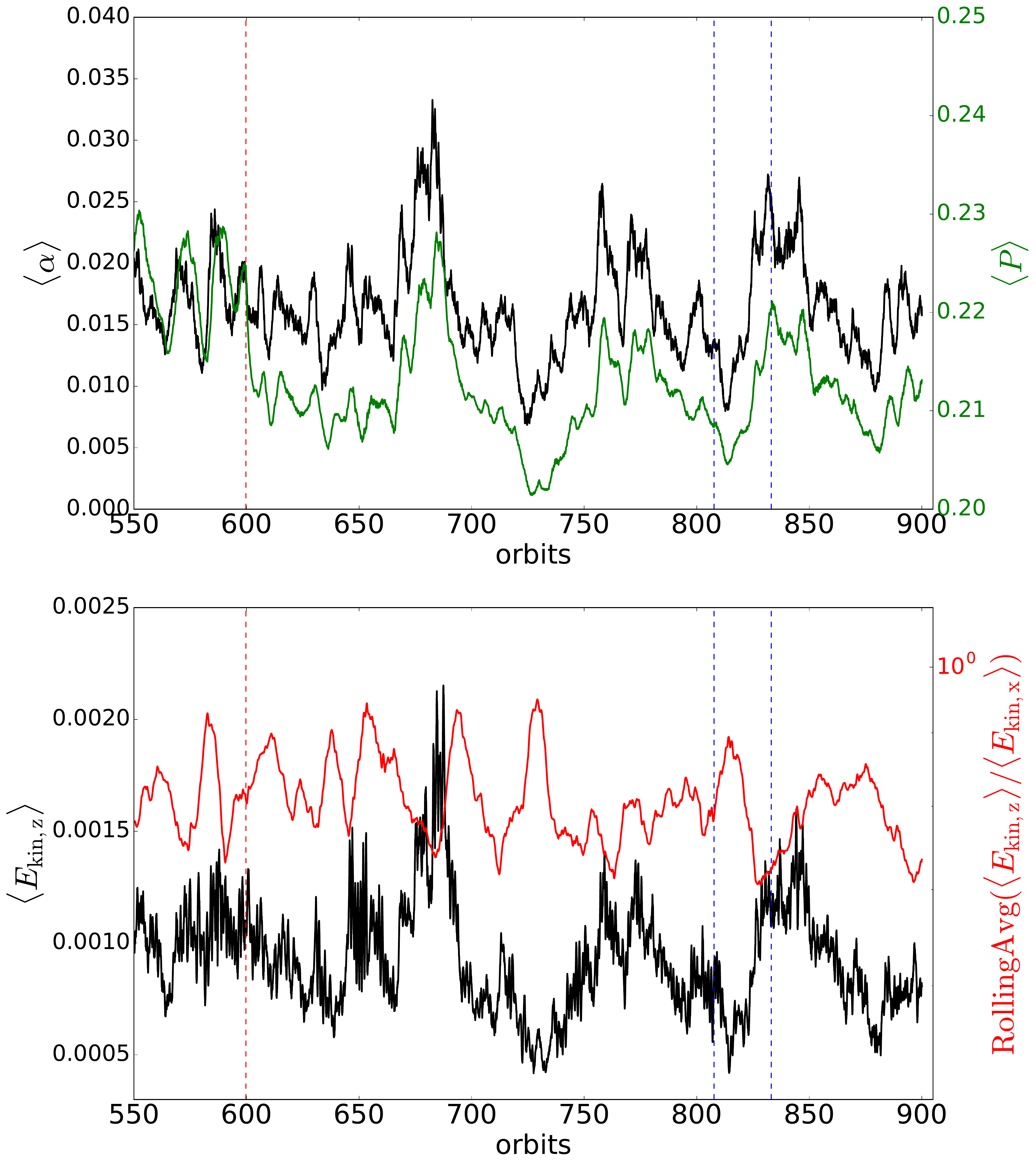}
\caption{Top: time-evolution of volume-averaged $\alpha$ (in black) and pressure (in green) from simulation NSTRMC44c6 ($\tau_\text{c} = 1$ orbit, $\eta = 10^{-4}$), exhibiting the outbursts characteristic of MRI/convective cycles. Bottom: time-evolution of vertical kinetic energy density (black) and a rolling average of the ratio of vertical to radial kinetic energy (red). The red dashed line indicates where the cooling timescale was lowered to $\tau_{\text{c}}= 1$ orbit from $\tau_{\text{c}} = 2$ orbits. The blue dashed lines at orbits 807 and 833 correspond to the times at which snapshots were taken during the convection- and MRI-dominated phases, respectively (see Figure \ref{FIGURE_MRISimsWithCoolingFlowFieldSnapshots2}).}
\label{FIGURE_NSTRMC44c6}
\end{figure}

\begin{figure*}
\centering
\subfloat[]{\includegraphics[scale=0.24]{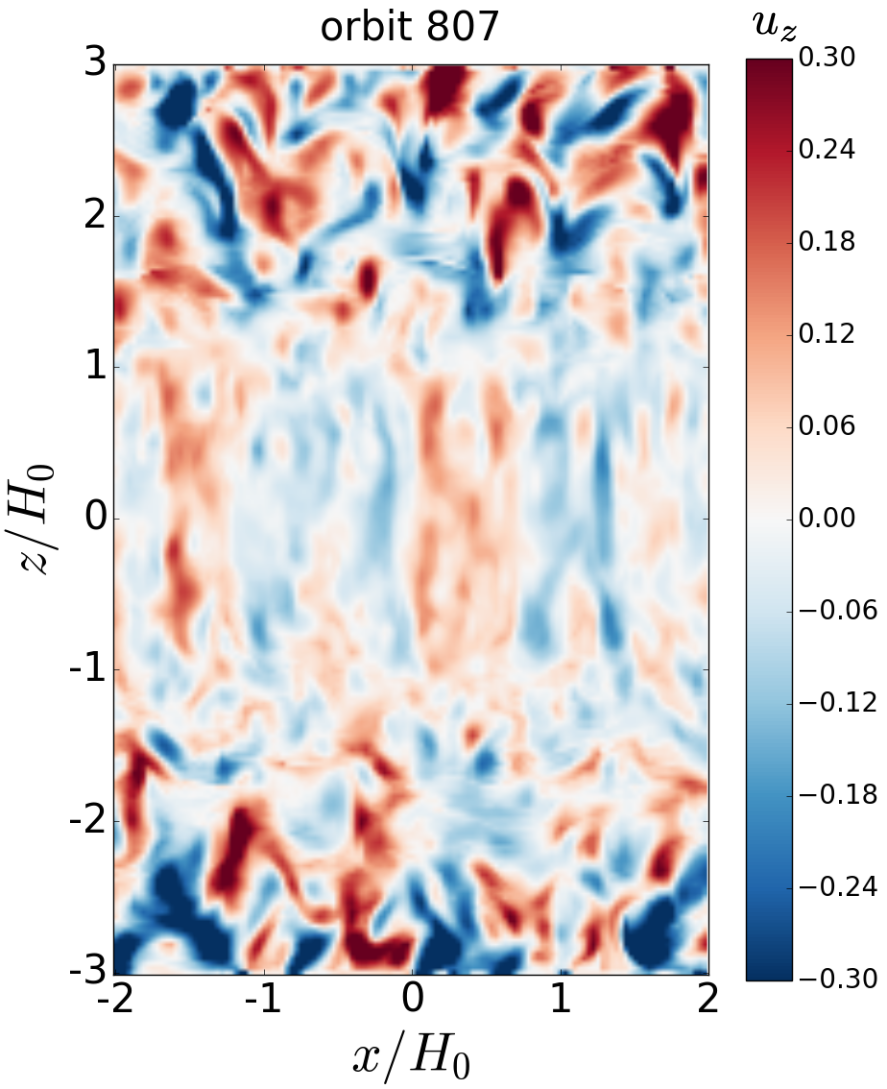}}\hspace{0.04em}
\subfloat[]{\includegraphics[scale=0.24]{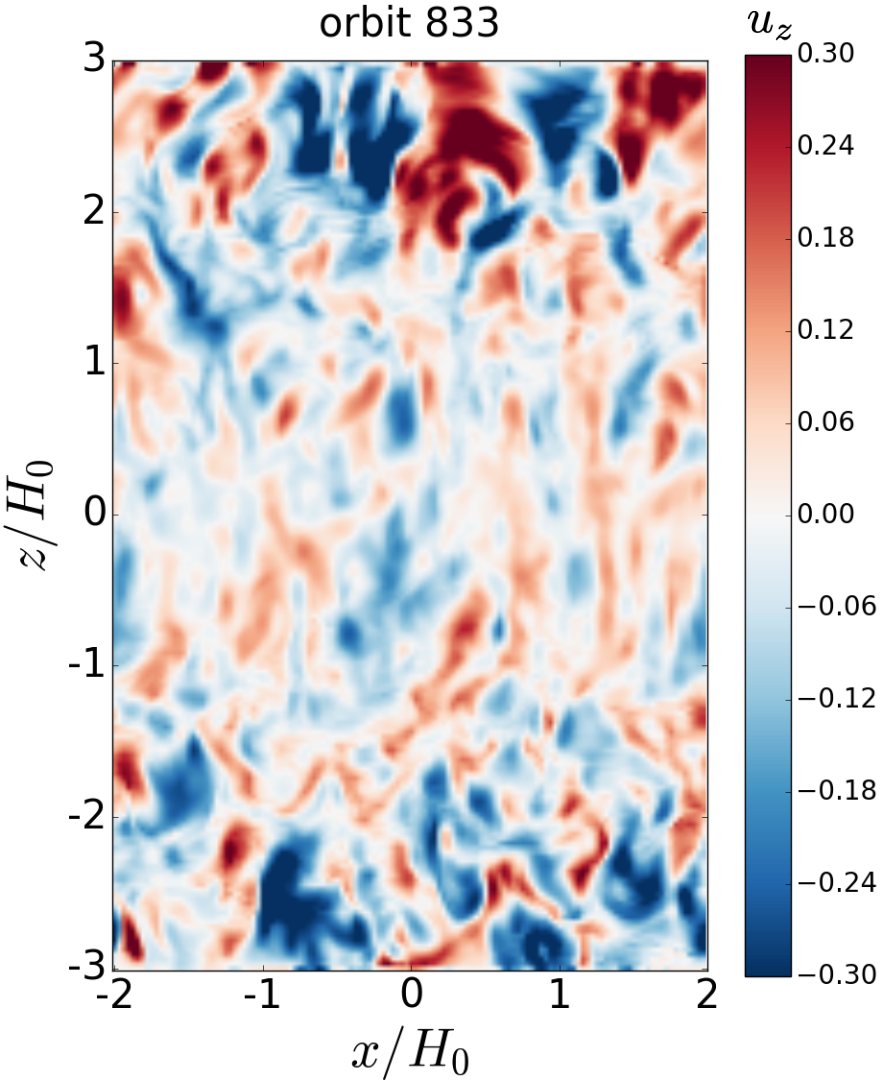}}\hspace{0.04em}
\caption{Colorplots of the vertical component of the velocity in the $xz$-plane taken from snapshots at different times from a simulation with a very low cooling timescale, which exhibited MRI/convective cycles (NSTRMC44c6). (a) convection-dominated phase, (b) MRI-dominated phase. The simulation employed a constant cooling timescale of $\tau_{\text{c}}=1$ orbit above $|z| = 0.75H_0$, and a uniform explicit resistivity of $\eta = 10^{-4}$.}
\label{FIGURE_MRISimsWithCoolingFlowFieldSnapshots2}
\end{figure*}

\subsection{Weak MRI/convective cycles}
\label{HEIGHTDEPCOOLING_WeakCycles}
The simulation described in the previous section exhibited clear MRI/convective cycles, with large enhancements in alpha and well-separated convection-dominated and MRI-dominated phases. However, it was also run at a relatively small magnetic Reynolds number (Rm $\sim 200$). To probe the parameter space at more realistic values of Rm of $10^4$ and $4\times10^3$, we ran a series of simulations at successively lower cooling timescales, with each simulation being restarted from the end state of the previous simulation (thus for each resistivity we move upwards along a vertical skewer in Figure \ref{FIGURE_ParameterSurvey}). In this new regime we recover solutions characteristic of MRI/convective cycles, but the amplitude of the cycles is smaller and the separation between the MRI and convection dominated phase is less clear than in the highly resistive runs. Therefore we refer to the solutions in this regime as \textit{weak} MRI/convective cycles.

An example of the typical behavior we observe as the cooling timescale is lowered (keeping the resistivity fixed) is given by the vertical skewer at $\eta=10^{-4}$ (simulations NSTRMC44c1 through NSTRMC44c7 in Table \ref{TABLE_MRISimulationsWithCooling}). As we decrease the cooling timescale, we move from a regime in which the simulations exhibit only MRI at $\tau_\text{c}=10$ orbits (e.g. simulation NSTRMC44c1) to a regime  characterized by MRI/convective cycles at $\tau_\text{c}=1$ orbit (simulation NSTRMC44c6). We observe correlated outbursts in the vertical kinetic energy density and in $\alpha$ (see Figure \ref{FIGURE_NSTRMC44c6}), the peaks of which we identify with the height of an MRI phase. The pressure (green curve) increases during outbursts as the disc heats up due to the dissipation of MRI turbulence, and drops during the convection-dominated phases (the following troughs). On the other hand, the convective phases can be positively identified by the peaks in ratio of vertical to radial KE (red curve in lower panel), which is anti-correlated with alpha: notice the peaks at $t=$695, 725, 810, and 875. (For clarity we have a plotted a rolling average of this ratio over $2.5\times10^{4}$ time-steps or about 8 orbits.) This anti-correlation mirrors that observed in our most extreme MRI/convective cycles simulations (see Figure \ref{FIGURE_MRISimsWithCoolingEnergies}). 
The main features of the flow field back up this interpretation, and we plot two examples in (see Figure \ref{FIGURE_MRISimsWithCoolingFlowFieldSnapshots2}). The left panel, taken from a putative convective phase, exhibits a large-scale pattern of updrafts and downdrafts, while the right panel, taken from an MRI phase, is more disordered. 

Overall the flow is quite turbulent (indeed in the first 100 orbits after the cooling timescale was lowered the cyclical pattern is difficult to discern). We therefore conclude that, rather than having two distinct MRI-dominated and convective dominated phases (as in our most resistive simulation), we are witnessing the two processes (MRI and convection) competing with one another. In this regime, the MRI is never fully suppressed by resistivity, and thus is somewhat free to impede convection. But the low cooling timescales force a sufficiently strong entropy gradient to enable convection for short periods of time, resulting in the intermittent bursts of convection we observe. 

Finally we have also run a simulation (NSTRMC44c7) at a very low cooling timescale of $\tau_\text{c} = 0.5$ orbits. In this case the cooling is so extreme, however, that the disc collapses and the MRI is extinguished.

\subsection{MRI-dominated}
\label{HEIGHTDEPCOOLING_MRIOnly}
For low resistivities  ($\eta < 5\times10^{-4}$) and moderate cooling timescale ($\tau_\text{c}=10$ orbits), we do not observe the MRI/convective cycles reported in the previous two sections. More importantly we also fail to find evidence that the MRI and convection `coexist', in the sense that they are continuously present at the same time. Rather we simply find MRI turbulence, though we do occasionally observe what appear to be convective bursts in the flow field, and these become more frequent in our more marginal simulations (e.g. at $\eta = 2.5\times10^{-4}$). 

An example of an MRI-dominated run is simulation NSTRMC44f1 (with parameters $\tau_\text{c}= 10$ orbits and $\eta = 10^{-5}$).\footnote{We have rerun this simulation with double the resolution (i.e. 64 cells per $H_0$) to ensure that the resistive scales were resolved; see NSTRMC44f1HR in Table \ref{TABLE_MRISimulationsWithCooling}.} The time-evolution of the kinetic and magnetic energy densities, and of the stresses, in the non-linear phase are similar to those measured in our box-cooled and uniformly cooled simulations with $\langle E_\text{kinz,z} \rangle \sim 10^{-3}$ and $\langle  E_\text{mag} \rangle \sim 10^{-2}$, respectively (see Appendix \ref{BOXCOOLING_TimeEvolutionOfAveragedQuantities}). While these quantities fluctuate with time in the non-linear phase, we observe no discernible bursts in the stresses as we do in our MRI/convective cycles runs. The time- and volume-averaged value of alpha is $\langle \langle \alpha \rangle \rangle \sim 0.015$ (compared to $\langle \langle \alpha \rangle \rangle \sim 0.014$ in our box-cooled simulation). Note that the simulations of \cite{hirose2014} and \cite{coleman2018} employ zero (explicit) resistivity, but slightly smaller thermal timescales of 5-10 orbits. Thus we have continued simulation NSTRMC44f1 for a further 100 orbits (NSTRMC44f2) but with a lower cooling timescale of 5 orbits. We find very similar time series to NSTRMC44f1 and do not detect any signs of convection.

\begin{figure*} 
\captionsetup[subfigure]{labelformat=empty}
\centering
\includegraphics[scale=0.35]{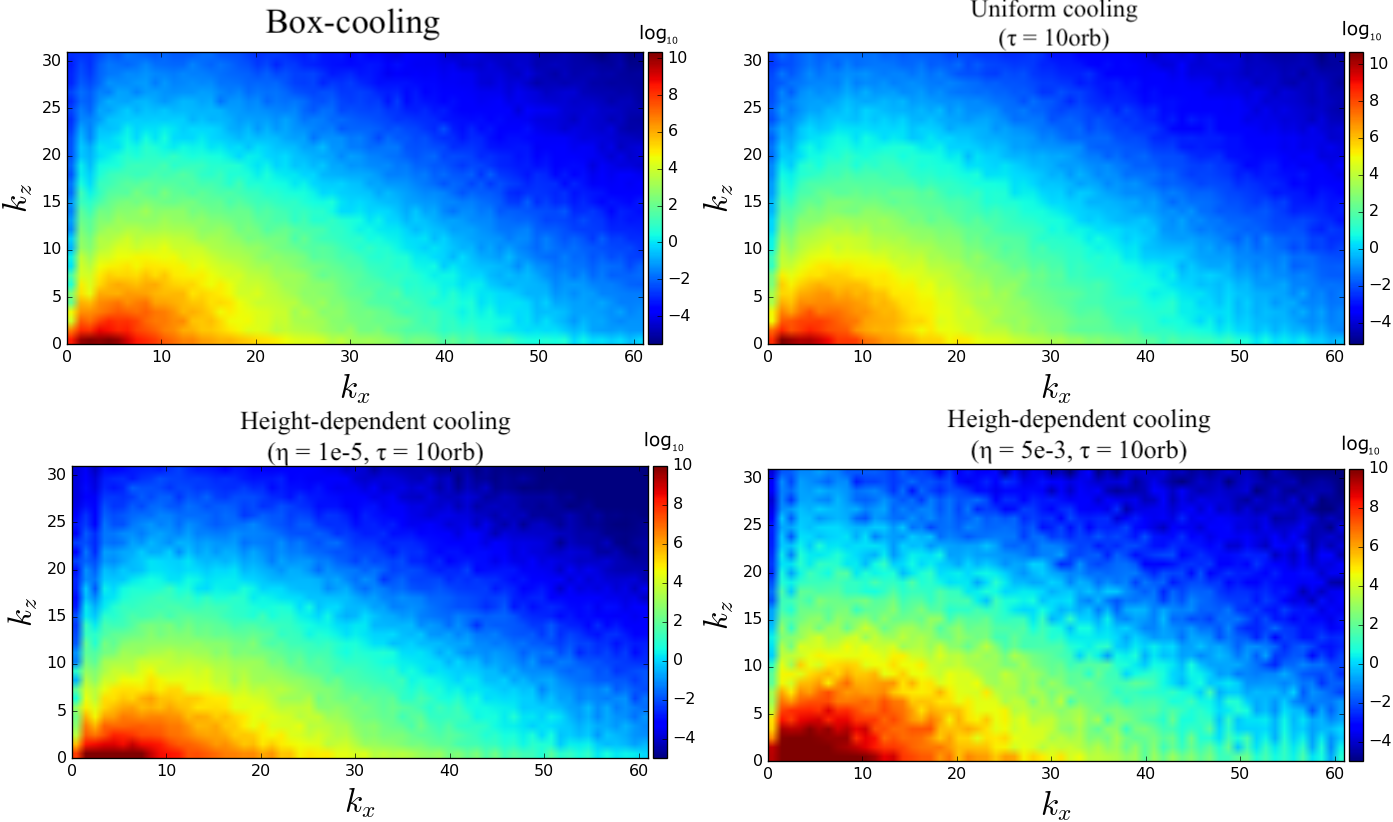}
\caption{Comparison of 2D specific kinetic energy spectra from simulations with different cooling prescriptions. See text for further details.}
\label{FIGURE_2DPowerSpectraCoolingPrescriptionsComparison}
\end{figure*}

NSTRMC44f1 (and also NSTRMC44f2) is an important example of the limitations of relying solely on the sign of the buoyancy frequency as a diagnostic for convection. The horizontal- and time-averaged vertical buoyancy frequency profile is clearly negative within around $\pm H_0$ of the mid-plane, reaching a minimum value of $\text{min}(\langle \langle N_B^2 \rangle_{xy} \rangle_t) \sim -0.14$. As we have already alluded to, however, this does not imply convection.

We now turn to the flow structure to discern any signal of convection in these runs. In Figure \ref{FIGURE_2DPowerSpectraCoolingPrescriptionsComparison} we plot the 2D power spectrum of the specific kinetic energy for our box-cooled simulation (top left), uniformly cooled simulation (top right), height-dependent cooling simulation (`straight MRI') with explicit resistivity $\eta = 10^{-5}$ (bottom left). We have also included the spectrum from the MRI-dominated phase of the height-dependent cooling simulation exhibiting strong MRI/convective cycles which was run with a resistivity of $\eta = 5\times10^{-3}$ (bottom right). The colorbars are logarithmic, and each panel corresponds to a separate simulation. 

The $y$-averaged spectra for the box-cooled, uniformly cooled, and highly resistive height-dependent cooled simulations are practically indistinguishable, and power is distributed more or less evenly along both the $k_x$ and $k_z$ directions. The y-averaged spectrum in the weakly resistive height-dependent cooling simulation (bottom left) is slightly flatter than the other simulations. However, the flatness is far less pronounced than in the spectrum captured during the convection-dominated phase of the MRI/convective cycles simulation discussed in the last section (see the top-hand panel of Figure \ref{FIGURE_2DPowerSpectra}), and the difference is marginal. Given the lack of evidence of any coherent vertical structure (e.g. convective plumes) in the weakly resistive run, it is difficult to attribute this feature to convection. It is more likely to be due to slightly different MRI properties in these different physical set-ups. 

\begin{figure}
\captionsetup[subfigure]{labelformat=empty}
\centering
\includegraphics[scale=0.45]{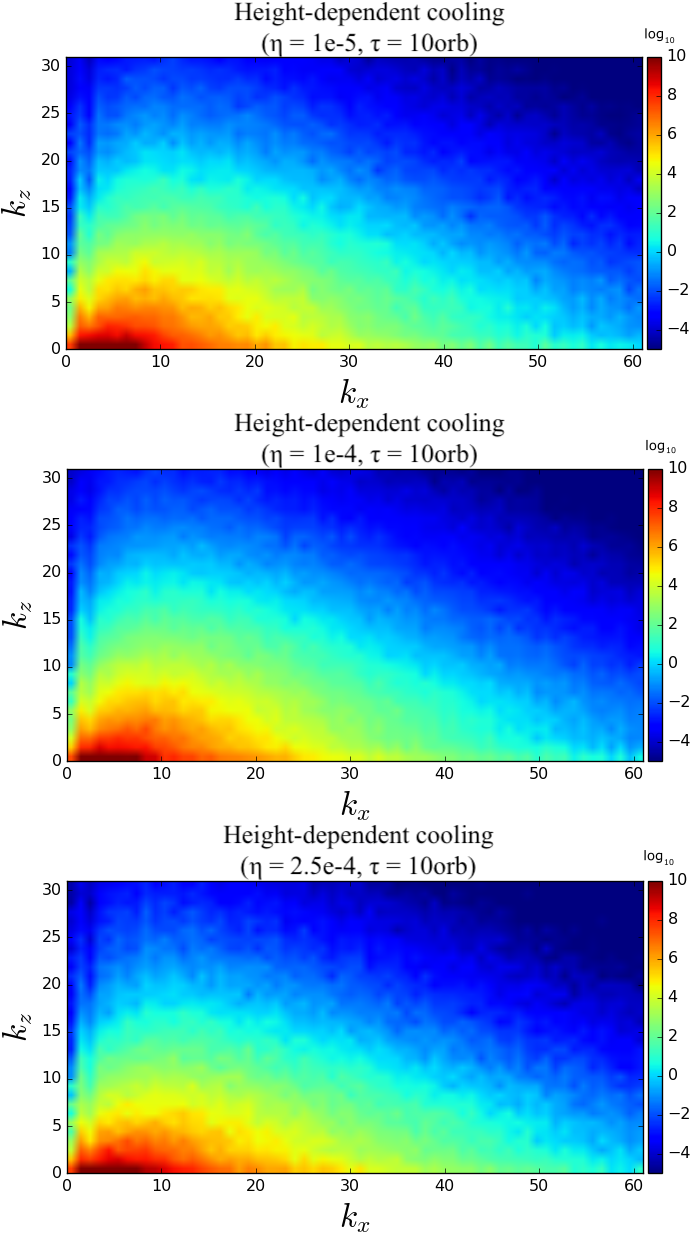}
\caption{2D power spectra of specific vertical kinetic energy from a simulations with cooling implemented above $|z| = 0.75H_0$. The colorbars are logarithmic. Each row corresponds to a separate simulation with height-dependent cooling, with the resistivity increasing from top to bottom, and all simulations employed a cooling timescale of $\tau_{\text{c}}=10$ orbits. For each spectrum the data has been taken within $\pm H_0$ of the mid-plane and averaged in time over several tens of orbits.}
\label{FIGURE_2DPowerSpectraHorizontalSkewer}
\end{figure}

Next we compare the spectra of our height-dependent cooling simulations at different $\eta$. Figure \ref{FIGURE_2DPowerSpectraHorizontalSkewer} compares the y-averaged 2D spectra from three simulations all run with a cooling timescale of $\tau_{\text{c}}=10$ orbits, but with resistivities of $\eta = 10^{-5}, 10^{-4}$, and $2.5\times10^{-4}$, thus moving along a horizontal skewer in the $(\tau_{\text{c}}, \eta)$ parameter space; see Figure \ref{FIGURE_ParameterSurvey}). There is very little difference between the spectra for the $\eta = 10^{-5}$ and $\eta = 10^{-4}$ runs, and both appear significantly more isotropic compared to the spectrum that we observed during the convective-phase of the simulation exhibiting MRI/convective cycles (see the top-hand panel of Figure \ref{FIGURE_2DPowerSpectra}), though still slightly more elongated in the $k_x$ direction than the box-cooled and uniformly cooled runs.

We have also tried looking at the 2D spectra of different quantities (vertical mass flux, vertical heat flux, density, and pressure) but we cannot detect any appreciable difference between the $\eta = 10^{-5}$ and $\eta = 10^{-4}$ simulations, nor enhanced structure in the vertical direction, for any of these quantities either. We have also looked at spectra computed from a fixed $y$-\textit{slice}, in case the $y$-average removed important non-axisymmetric features. Using these new spectra to compare the different runs, we also failed to turn up any meaningful difference. 

\section{Discussion}
\label{DISCUSSION}
We found that vertically stratified simulations of the MRI with a perfect gas equation of state, a height-dependent explicit cooling prescription, and uniform explicit resistivity can lead to convective/MRI cycles. Several questions naturally arise. To what extent are the MRI bursts (i.e. enhancements in $\alpha$) actually related to convection? Furthermore, if the outbursts are related to convection, to what extent can we find support for the hypothesis presented in \cite{hirose2014} that convection can `seed'  net-vertical flux MRI? Can narrow boxes (in radius) mitigate the large-scale convective cells (which typically have a width $\sim H_0$) observed in the convective/MRI cycles simulations? Finally, what are the applications to dwarf novae and to previous work on MHD convection in discs? These questions are discussed in the following subsections.

\subsection{Effect of resistivity}
\label{DISCUSSION_EffectOfResistivity}
\cite{simon2011} (SHB2011) carried out vertically stratified \textit{isothermal} MRI simulations with explicit resistivity and viscosity, and found that at certain magnetic Prandtl numbers Pm (where Pm$\, \equiv \nu/\eta$) accumulation of toroidal field could switch the MRI back on again after it was initially killed by the resistivity. To what extent are the results of our MRI/convective cycles simulation a manifestation of this phenomenon? To investigate this, we reran the simulation described in Section \ref{HEIGHTDEPCOOLING_StrongCycles} \textit{without} any explicit cooling, but otherwise with exactly the same set-up, including an explicit uniform resistivity of $\eta = 5\times10^{-3}$ (see simulation NSTRMC44e1NoCool in Table \ref{TABLE_MRISimulationsWithCooling}). 

\begin{figure}
\centering
\captionsetup[subfigure]{labelformat=empty}
\subfloat[]{\includegraphics[scale=0.21]{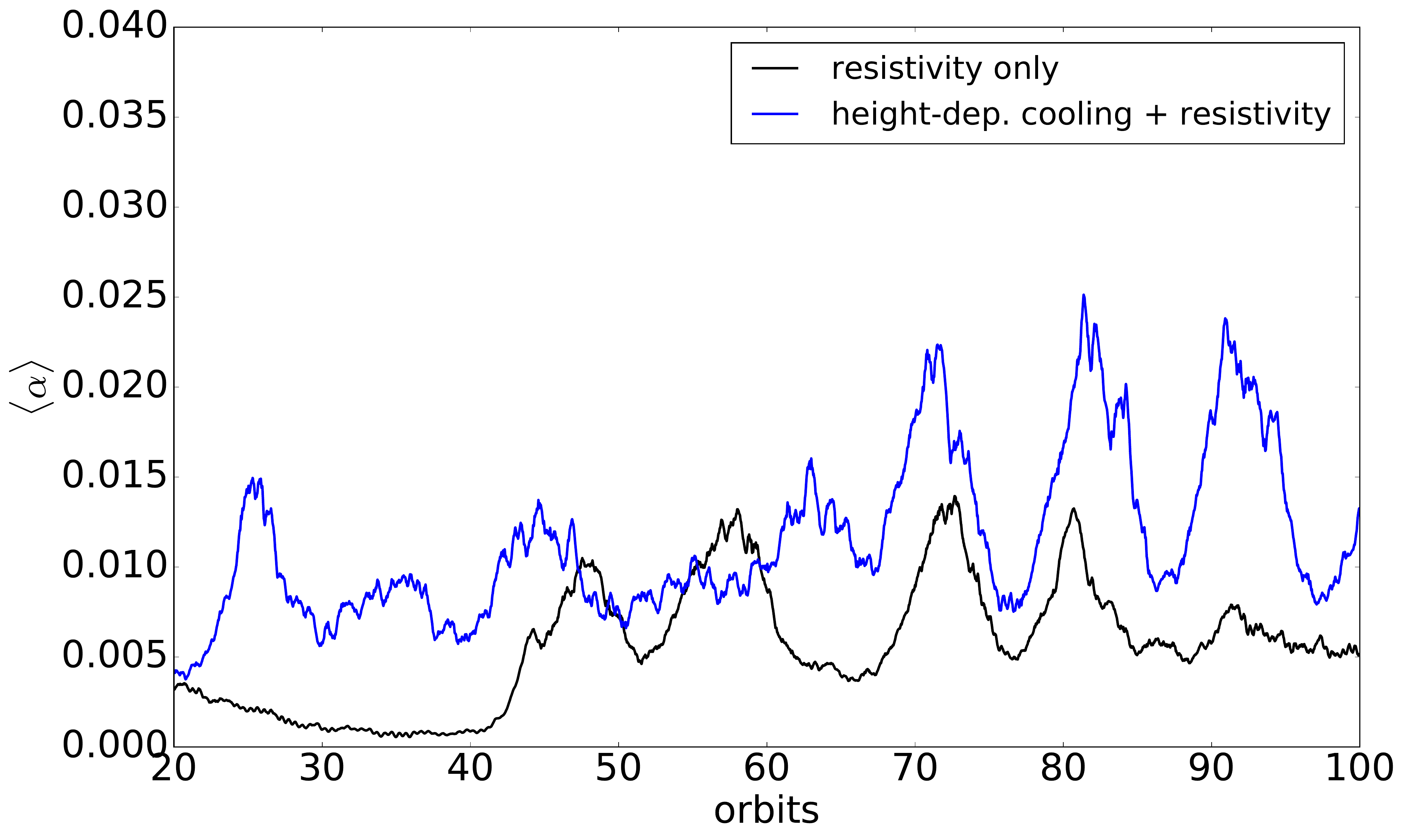}}
\caption{Time-evolution of volume-averaged alpha viscosity from a simulation run with height-dependent cooling and explicit resistivity with ($\tau_{\text{c}}=10$ orbits, $\eta = 5\times10^{-4}$) (blue curve), and from a simulation run with the same explicit resistivity but no cooling (black curve).}
\label{FIGURE_NSTRMC1044b1Comparison}
\end{figure}

We find that vertical kinetic energy density drops by an order of magnitude immediately after initialization (from $\langle E_{\text{kin},z} \rangle \sim 5\times10^{-3}$ to $\sim 10^{-4}$), as does the magnetic energy density (from $\langle E_{\text{mag}} \rangle \sim 3\times10^{-2}$ to $\sim 4\times10^{-3}$). Thereafter the kinetic energy continues to decrease steadily (though fluctuating) over the duration of the simulation (100 orbits). There is a gradual increase in the magnetic energy, which saturates at around $10^{-2}$ by the end of the simulation. There is a steep drop in both the Reynolds and Maxwell stresses after initialization and both remain of order $10^{-5}$ for the duration of the simulation. In addition the flow field appears laminar and there are neither visible signs of convection, nor of MRI turbulence. We conclude that in the absence of convection the explicit resistivity kills the MRI which is not reseeded by any other process for the duration of the simulation. Thus we can confirm that, at least for the largest resistivity we investigate, the outbursts and MRI/convective cycles are not being driven by resistivity and bear no resemblance to the dynamics uncovered by SHB2011. On the other hand, we have the remarkable result that convection in this strongly resistive environment actually helps the MRI persist to lower Rm then it would otherwise do. 

At smaller resistivities the picture is somewhat more complicated. In Figure \ref{FIGURE_NSTRMC1044b1Comparison} we plot the time-evolution of $\alpha$ (in blue) from a simulation (NSTRMC44b1) with both height-dependent cooling (with $\tau_\textrm{c} = 10$ orbits) an explicit resistivity of $\eta = 5\times10^{-4}$, which is an order of magnitude smaller than that that used in NSTRMC44e1. This simulation also exhibits clear bursts in $\alpha$, though the peaks are smaller ($\alpha \sim 0.015$-$0.025$, about half the size of the peaks in $\alpha$ observed in the $\eta = 5\times10^{-3}$ run), and spaced closer together. Other than that the results are broadly similar to those in NSTRMC44e1, and we observe both large scale convective cells between the bursts in $\alpha$ and MRI turbulence during the bursts. One major difference however, is the behavior of this simulation when the cooling is turned off. We have rerun NSTRMC44b1 with exactly the same set-up, but without cooling (simulation NSTRMC44b1NoCool in Table \ref{TABLE_MRISimulationsWithCooling}), and plotted the time-evolution of $\alpha$ from this run as a black curve in Figure \ref{FIGURE_NSTRMC1044b1Comparison}. The buoyancy frequency is positive everywhere in this run, and thus there is no convection. Unlike earlier, this lower resistive run can sustain the MRI in some form, and it is somewhat bursty in $\alpha$, though these bursts are smaller in magnitude than those in the simulation with cooling. 

It is tempting to draw parallels between this (resistive only) simulation and some of the vertically stratified isothermal resistive simulations of SHB2011 (in particular see the red curve on the right-hand panel of their Figure 10). Thus NSTRMC44b1 without cooling might be the perfect gas analogue of what SHB2011 found in their isothermal simulations, where accumulation of toroidal magnetic field, rather than convection, was reseeding the MRI. This comparison should be made with caution, however, as the magnetic Reynolds number is smaller in our simulation (Rm = 2000) than in the aforementioned simulation of SHB2011 (where Rm = 3200). In addition SHB2011 varied the explicit viscosity (and thus the magnetic Prandtl number) in their simulations rather than the resistivity. Finally the time scales of the bursts are very different between the two simulations. In the simulations of SHB2011 the stress builds up and then decreases again over several hundred orbits, whereas the bursts observed in our resistive-only simulation recur much more rapidly, on timescales of just 10-15 orbits. In any case, while it appears that resistivity is contributing to the bursty behavior in this simulation, the fact that the bursts are consistently larger when height-dependent cooling is added seems to suggest that convection can enhance this behavior.

\subsection{Is convection seeding net-vertical-flux MRI?}
\label{DISCUSSION_ConvectionSeedingMRI}
This brings us to our second question: is convection potentially reseeding net-vertical-flux MRI leading to an enhancement in alpha as claimed by \cite{hirose2014}? \cite{hirose2014} hypothesize that vertical convective motions drag magnetic field lines upwards, thus creating net vertical field that seeds the (powerful) net-vertical field MRI \citep[see][]{hawleygammiebalbus1995}. The net-vertical flux MRI is characterized by radial \textit{channel flows} during its linear phase. It is unlikely, however, that channel flows are being seeded in the simulations given the short radial wavelength of the $B_z$ fluctuations, and indeed visual inspection of the flow field (i.e. $B_x$ and $B_y$ in the $xz$-plane) reveals no streaky motions (in $x$) characteristic of these channel flows. In addition we find no noticeable difference in $B_x$ and $B_y$ at the onset of an outburst compared to the middle of an outburst. 
Note, however, that having stronger vertical magnetic field, even if it varies with $x$, still aids in seeding (possibly ZNF) linear MRI, even if doesn't quite generate the channel modes associated with net vertical flux version.

\begin{figure}
\centering
\captionsetup[subfigure]{labelformat=empty}
\subfloat[]{\includegraphics[scale=0.22]{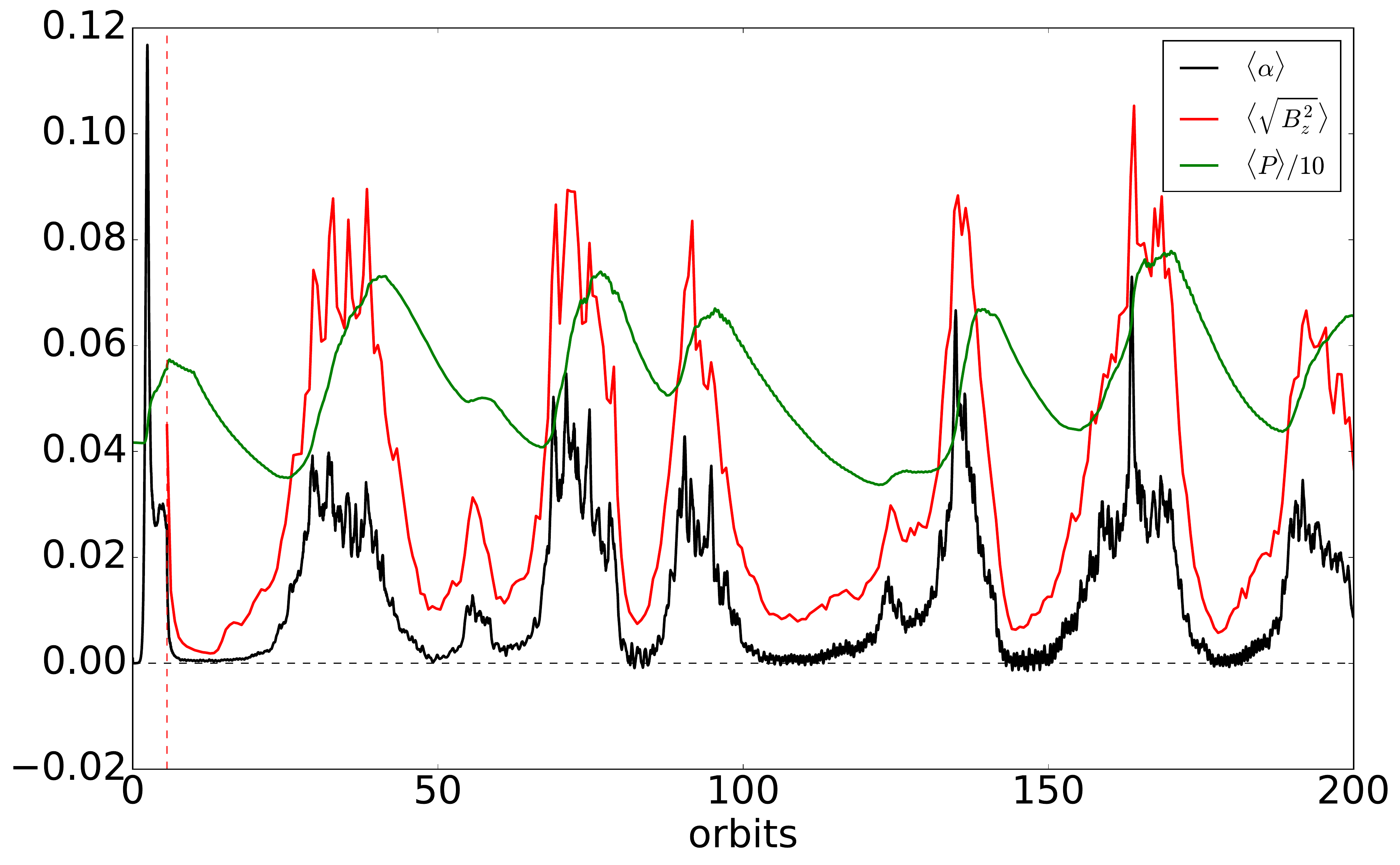}}
\caption{Time-evolution of volume-averaged alpha viscosity (black), root-mean-square vertical component of the magnetic field (red), and pressure divided by 10 (green) from a simulation exhibiting MRI/convective cycles (NSTRMC44e1) with ($\tau_{\text{c}}=10$ orbits, $\eta = 5\times10^{-3}$).}
\label{FIGURE_AlphaBx3rms}
\end{figure}

More revealing is the time-evolution of the root-mean-square vertical magnetic field component $\langle B_z^2\rangle^{1/2} $ (see Figure \ref{FIGURE_AlphaBx3rms}). This exhibits a clear phase shift compared to the time-evolution of $\langle \alpha \rangle$: the vertical magnetic field appears to increase just before $\langle \alpha \rangle$ does, which might be indicative that convection is building up vertical field just before an outburst. The lag between the rms vertical magnetic field and the stress is approximately a few orbits, which is comparable to the MRI growth time. To quantify this, we have measured the growth rates at the onsets of the outbursts by fitting the slopes of the magnetic energy density (cf. middle panel of Figure \ref{FIGURE_MRISimsWithCoolingEnergies}): the growth rate of the bursts during the non-linear phase is typically $\sim 0.02\,\Omega$. This is an order of magnitude less than the growth rate $\sim 0.4\,\Omega$ expected from net vertical flux (NVF) resistive MRI.\footnote{We have calculated this by solving the dispersion relation for resistive NVF MRI given by Equ 25 in \cite{pessah2008}.} This shows that what we have here is, as suspected, not true linear net vertical flux MRI, even if there is some increase in the net vertical flux prior to a burst, but a weakened version thereof due to the highly variable $B_z$ in the radial and azimuthal directions.

It should also be mentioned that, quite aside from the generation of a mean $B_z$, it has long been established that convection can itself work as a turbulent dynamo, producing disordered magnetic fields up to equipartition strengths (see review by \cite{rincon2019} and references therein). The zero net flux MRI might be able to use this dynamo field as a seed, though we don't pursue this angle explicitly, concentrating on the more straightforward and probably more dominant effect of the large-scale $B_z$ generated.

\subsection{Dependence on radial box size}
\label{DISCUSSION_DependenceOnBoxSize}
All the flux-limited-diffusion simulations exhibiting convective/radiative cycles in \cite{hirose2014} were carried out in very narrow boxes ($H_0$ in the radial direction). The disadvantage of narrower boxes is that they limit the radial sizes that convective (or other motions) can exhibit. This might particularly be true in the case of the large-scale convective cells observed in NSTRMC44e1: the typical size of the large-scale convective eddies in this run was found to be $\sim1.3H_0$ (cf. right-hand panel of Figure \ref{FIGURE_MRISimsWithCoolingFlowFieldSnapshots}). Thus we briefly investigate the dependence of the results of this simulation on the radial box size by comparing the results of NSTRMC44e1 ($L_x = 4H_0$) with NSTRMC46 ($L_x = 2H_0$) and NSTRMC45a3 ($L_x = H_0$), keeping all other parameters the same.

As the radial box size is reduced, we find that the onset of the first outburst is delayed: from around orbit 25 (at $L_x = 4H_0$) to around orbit 40 (at $L_x = 2H_0$) up to orbit 60 (at $L_x=H_0$). Consequently we find also find that the number of outbursts is reduced as the radial box size decreases. We observed 3 outbursts during the first 100 orbits in the $L_x = 4H_0$ simulation, 2 outbursts in the $L_x = 2H_0$ simulation, and just one outburst in the $L_x = H_0$ simulation.  The magnitude of the outbursts (measured in terms of the maximum value of $\langle \alpha \rangle$ measured during the outbursts) is about the same at all three radial box sizes. Crucially, large-scale convective cells are observed in the flow field only in the $4H_0$ simulation: the $2H_0$ and $H_0$ simulations exhibit only narrow convective cells during their low stress phases. Thus it appears that a radial box sizes of $2H_0$ and $H_0$ are insufficient to observe the large-scale convective cells.

\subsection{Comparison with recent DNe/AM CVn simulations}
\label{DISCUSSION_ApplicationsToDwarfNovae}
The convective/radiative cycles observed in the radiative MHD simulations of \cite{hirose2014} (HBK2014) have outbursts that recur on timescales similar to those in our MRI/convective cycles runs. However, besides the more complicated radiative transfer physics, which we mock up using a simple height-dependent cooling prescription, there are appreciable differences between our results and those of HBK2014, and they cannot be regarded as issuing from the same cause. The HBK2014 cycles appear to be driven by the opacity's temperature and pressure dependencies, and manifest as alternating phases in which either the advective heat flux dominates or the radiative heat flux dominates. Our height-dependent cooling simulations are always advection dominated (whether the flux is associated with the MRI or actual convection). Furthermore, convection appears to be present \textit{during} the outbursts in their simulations, whereas we find the opposite in our runs, i.e. that convection is suppressed during outbursts (when MRI turbulence takes over), but dominant during the periods of low stress.

\cite{coleman2018} carried out local, radiative MHD simulations very similar to those of HBK2014, but employed opacity and equation of state tables relevant to the helium-rich AM CVn discs. A key difference in most of their simulations compared to those of HBK2014 is that the vertical heat transport is persistently dominated by advection (or convection, according to their criteria), an outcome they attribute to the system lying between the two opacity maxima resulting from the two ionisations of helium at $\sim 3\times10^{4}$ K and $\sim 7\times10^{4}$ K, respectively. Thus these results might provide a better point of comparison to ours. 

Three different outcomes (each corresponding to a different initial surface density) are shown in Figure 7 of \cite{coleman2018}. The top panel shows `persistent convection', the middle panel shows `mostly persistent convection',  and the bottom panel shows `intermittent convection' (i.e. convective/radiative cycles as see in HBK2014). Both the top and and the middle cases have a ratio of advective to total heat flux close to unity for the duration of the simulations, and are thus similar in that respect to our simulations which are always completely advection dominated. In particular note that in the middle panel (the `mostly persistent convection' case), the `advective' Mach number (cf. Equation 4 in their paper) is mostly anti-correlated with $\alpha$, and that $\alpha$ is quite bursty. This behavior is reminiscent of the simulation NSTRMC44c6 discussed in Section \ref{HEIGHTDEPCOOLING_WeakCycles} in which $\alpha$ was very bursty, the ratio of vertical to radial KE was largely out of phase with $\alpha$, and in which we observed frequent bursts of convection. Indeed the opacity in their run is mostly clustered around the second Helium ionisation maximum (see Figure 6 of their paper). The large opacity likely leads to a steeper entropy gradient which is more favorable for convection to dominate over the MRI. The top panel of their Figure 7, on the other hand, exhibits less bursty behavior and the advective Mach number is largely in phase with $\alpha$. This is similar to the behavior of our simulations with small resistivity and long ($\tau_\text{c} \sim 5$-$10$ orbits) cooling timescales, in which we found the flow was dominated by MRI turbulence rather than convection. Indeed in their run the opacity is largely concentrated in the trough between the two opacity maxima, and the lower opacity likely leads to a shallower entropy gradient, which is less favorable to convection. Thus what \cite{coleman2018} refer to as `convection' in this run might really just be MRI turbulence. 

Apart from the time variable behavior just described, note that \cite{coleman2018} find that $\alpha \gtrsim 0.1$ for the duration of their simulations (in all three cases). As they point out, this calls into question the role of the cycles observed in HBK2014 in enhancing $\alpha$, but it is also significantly higher than the values we find in our simulations (we have seen $\alpha \sim 0.08-0.1$ only during the very peaks of the MRI bursts in our most resistive simulations).

A further difference between our simulations and those of HBK2014 \citep[and of][]{coleman2018} is that their simulations are ideal, whereas we employ an explicit resistivity. \cite{scepi2018a}, however, did carry out radiative (zero net flux) MHD simulations with explicit resistivity, though they employ a time-dependent resistivity whereas we keep the resistivity uniform in space and time. In a run in which the magnetic Reynolds number within the mid-plane was around Rm$\, \sim 2\times10^{4}$, they did not observe radiative/convective cycles. This is consistent with our run NSTRMC44c1 (with $\textrm{Rm} \sim 10^{4}$) in which we also do not find any cyclical behaviour. At lower magnetic Reynolds numbers  ($\textrm{Rm} < 10^{4}$), however, \cite{scepi2018a} find that the MRI dies, whereas we observe convection which reseeds the MRI, leading to the aforementioned MRI/convective cycles. \cite{fleming2000} find that in their simulations with net \textit{vertical} flux there is cyclical revival of the MRI down to values as low as $\textrm{Rm} \sim 50$,\footnote{Note that the simulations of \cite{fleming2000} were unstratified.} though \citep{scepi2018b} carried out their own net vertical flux simulations and find this revival does not occur when time-dependent resistivity is used. It is possible, therefore that the use of a time-dependent resistivity would suppress the revival of the MRI by convection at low $\textrm{Rm}$ in our ZNF simulations, too. On the other hand the resistive simulations of \cite{scepi2018a} do not have convection since they lie on the optically thin lower branch.

\subsection{Application to dwarf novae}

Finally, we wish to comment on the relevance of our results to dwarf novae. With respect to the resistivity, \cite{gammie1998} have estimated that the magnetic Reynolds number in the U Gem type dwarf nova SS Cyg in quiescence is around Rm $\sim 10^{3}$. This falls squarely within the regime in which we observe MRI/convective cycles and powerful MRI outbursts: for example, our run NSTRMC44b1 had Rm = $2\times10^{3}$ (see Figure \ref{FIGURE_NSTRMC1044b1Comparison}). An interesting corollary of our results is that we have found that convection appears to prolong MRI activity to lower Rm than the critical value of Rm$\,\sim 10^{4}$ quoted by \cite{gammie1998} and found in previous ZNF simulations \citep{fleming2000,scepi2018a}. We caution however that in dwarf novae the lower branch is optically thin and thus convection will be impeded, a set-up not captured in in our idealized simulations.

Resistivity is only part of the picture, however, and during the quiescent phase of dwarf novae other non-ideal MHD effects, such the Hall effect, might also play a role \citep{coleman2016}, though they are likely subdominant compared to Ohmic resistivity (\cite{scepi2018a}; see also \cite{coleman2018} for a brief discussion). 

Finally, we note that the shortest recurrence time between dwarf novae outbursts is around 7 days \citep{warner1995}, whereas the interval between successive MRI-dominated outbursts in our most resistive simulations is around 50 dynamical timescales. For a disc of size $5\times10^{-3}\,$AU around a white dwarf of mass $1\,M_{\odot}$ the outermost annulus will execute 50 orbits in about 6.4 days, so most of the disc will undergo at least one convective/MRI cycle during its period of quiescence, with the gas near the inner radii possibly executing several tens of cycles. The connection of these oscillations to any observed variability must take into account if the cycles organise themselves into coherent global structures, and how that might impact on accretion. This is left to future work.

\section{Conclusions}
\label{CONCLUSIONS}
Motivated by our findings in \cite{heldlatter2018} that hydrodynamic convection can transport angular momentum outward in discs, that the MRI can act as a heat source for convection, and by claims from radiation MHD simulations that angular momentum transport can be enhanced when the two instabilities interact in dwarf novae \citep{bodo2012, hirose2014}, we have undertaken a study of the interplay between convection and the MRI through controlled, three-dimensional, vertically stratified, zero-net-flux (ZNF), fully-compressible MHD simulations in \textsc{PLUTO}.

Previous work on the topic has used the negativity of the buoyancy frequency to determine whether convection is present or not \citep{hirose2014}. We have discussed how in a turbulent fluid this is only a necessary condition, not a sufficient one, because the turbulent transfer of heat and momentum by the MRI leads to an effective thermal diffusivity and viscosity, and therefore an effective Rayleigh number. MRI turbulence might limit the onset of convection by lowering the effective Rayleigh number of the flow below the critical value at which the unstable entropy gradient can overcome the diffusive effects of the turbulent diffusivities (though this does assume that turbulent transport coefficients behave in a diffusive manner). Furthermore we have argued that the two instabilities if both present are unlikely to be additive, and that we expect either one or the other to dominate the flow at any given time. At the very least, the two should interact in non-trivial ways: the MRI provides residual heating that can, given sufficient cooling, set up the unstable entropy gradient that is necessary for convection; the disordered motions associated with each instability work on similar scales and thus can impede the operation of the other; and yet we find in some simulations that convection can help reseed the MRI by producing vertical magnetic field. Distinguishing between the two instabilities in a turbulent flow field remains a challenge, however, and a key aim of future work should be to develop better diagnostics for doing so.

We carried out inviscid numerical simulations without any explicit cooling ('box-cooled' runs; see appendices), though we still allowed for advection of mass and energy across the vertical boundaries in order to prevent the disc from heating up and expanding until it filled the box, as was the case in the simulations of \cite{bodo2012}. One might assume that these simulations would provide a favourable environment for the onset of convection, because the only way to remove heat from the midplane is via advection. However, the system never exhibits an unstable entropy gradient, and the buoyancy frequency is positive everywhere in the disc, thus ruling out convection. However, we also find that box-cooled runs (and also runs with uniform explicit cooling) are plagued with pathologies that make determining the true vertical structure due to the MRI alone difficult.

To dispense with the pathologies associated with box-cooled and uniformly cooled runs, and to obtain a negative entropy gradient in MRI simulations (and therefore ensure that at least a necessary condition for convection is satisfied), we implemented height-dependent explicit cooling that switched on only above $|z|=0.75H_0$. We also introduced an explicit resistivity. By adjusting the cooling timescale we could control the strength of the unstable entropy gradient, and by adjusting the resistivity we could control the strength of the MRI. We find that convection and the MRI tend not to coexist, or at least do not interact in an additive manner. 

First, at low resistivities ($\eta < 2.5\times10^{-4}$) and moderate cooling times, the MRI is too strong and disrupts any convective modes before they can become coherent plumes. Despite the negativity of the buoyancy frequency, the results look similar to those of our convectively stable box-cooled simulations. We provide diagnostics of the flow's spatial spectra to illustrate this point.

At higher resistivities ($\eta \geq 5\times10^{-4}$), significant enough to impede the MRI, the picture is more complicated. We observe `convective/MRI cycles' in which there are alternating convection-dominated and MRI-dominated phases. Furthermore, we observed the same large-scale convective cells that we found in our hydrodynamic simulations \citep[see][]{heldlatter2018}, demonstrating that these structures are robust when the heat source is implemented in a self-consistent manner via residual heating from the MRI (or gravitational contraction), rather than through a heat source that was put in by hand, as in our hydro simulations. At the end of a convective cycle, the convection appears to reseed the MRI via the production of vertical field, leading to very strong MRI bursts in which $\alpha$ can reach as high as $\alpha \sim 0.08$. While we cannot find any evidence of the strong radial `channel modes' associated with the net vertical flux MRI prior to these outbursts, we do see a clear increase in the vertical rms magnetic field prior to each outburst. Lastly, when we remove the cooling prescription (and hence convection) in these very resistive runs, the MRI just dies away - and so we have the remarkable result that convection actually helps the MRI persist to lower Rm than it would otherwise do.

At low resistivities and low cooling times, the MRI is not impeded but the entropy gradient is very strong. In this regime, we also observe alternating convective and MRI phases, though the temporal behaviour is more chaotic, disordered, and bursty. Unlike the high $\eta$ runs, the MRI is never fully suppressed and we witness the two instabilities compete, with convection intermittently breaking through. We term this regime `weak MRI/convective cycles'.

Though significantly idealised, our simulations provide a template to help understand the complex fluid dynamics taking place when convection and MRI turbulence feature, and should provide a means to interpret the complicated data generated by simulations involving more realistic radiative physics. Our results are not comparable to the advective/radiative cycles discovered by \citet{hirose2014}, as our simulations are always `advective'. However, they may elucidate the behaviour shown in \citet{coleman2018}: their `persistent convection' cases may map on to our `straight MRI' regime, in which the MRI turbulence performs all the vertical heat transport, while their `mostly persistent' convective runs may correspond to our runs displaying weak MRI/convective cycles. Future work dedicated to these links should be able to determine how far the analogies can go.  Finally, though our focus has been on DNe, we expect an equally vexed relationship between the MRI and convection in very hot accretion flows, where (quite notably) convection has been difficult to demonstrate \citep[e.g.][]{hawley2001magnetohydrodynamic,hawleybalbus2002}. Added complications here include the fact that the unstable entropy gradient is both radial and vertical, with the MRI capable of extracting energy from the former \citep{balbus2002nature}, and the weakly collisional nature of the plasma, which permits a wider range of instability because of anisotropic heat conduction \citep{balbus2000stability,balbus2001convective}

\section*{Acknowledgements}
This work was funded by a Science and Technologies Facilities Council
(STFC) studentship. The authors acknowledge useful input from Geoffroy Lesur, Nicolas Scepi, Jim Stone, Matt Coleman, and Hubert Klahr. LEH would like to thank William B{\'e}thune for stimulating conversations and advice on using the PLUTO code.

\section*{Data availability}
The data underlying this article will be shared on a reasonable request to the corresponding author.

\bibliographystyle{mnras}
\bibliography{2020LEHMHDConvectionPaper_Bibliography} % if your bibtex file is called example.bib

\appendix

\section{Simulations without explicit cooling}
\label{BOXCOOLING}
\subsection{Motivation}
\label{BOXCOOLING_Motivation}
In this section we describe MRI turbulence with no explicit cooling (and no explicit resistivity). This set-up is simpler than the height-dependent  cooling prescription employed in our resistive runs (see Section \ref{HEIGHTDEPCOOLING}). The idea here is that without a cooling function the only way a thermal balance is obtained is via losses from the vertical boundaries (thus we refer to this set-up as \textit{box cooling}). While it is true that this means the saturated state depends on the box size and boundary conditions, this set-up presents very `pure' conditions for convection to develop: heat generated near the midplane must somehow get to the boundaries, and we might expect convection to arise in order to do the job. As we show, this is \emph{not} what happens: instead we find the disc is convectively stable. Furthermore, we observe a large outward heat flux. This appears to be driven not by MRI turbulent motions, but as a consequence of a mean vertical flow which is set up by the combination of advection of fluid across the vertical boundaries and displacement of fluid inside the box by fluid added by the mass source term. While volume-averaged quantities (such as $\alpha$) do not appear particularly sensitive to this numerical pathology, it has marked consequences for the vertical structure of the disc, as we show below. We believe that the demonstration is nevertheless useful and informative.

For the sake of brevity, we mostly restrict our discussion in the following three sections to a single (fiducial) simulation (NSTRMRI4c). We have repeated this run at a higher resolution of $256\times256\times392$ which is equivalent to $64$ cells per scale-height (cf. NSTRMRI4d in Table \ref{TABLE_MRISimulationsWithoutCoolingDependenceOnBoxSize}), but the results were the same as those for the lower resolution run and so we omit further discussion of that run for brevity. In addition, we have repeated our fiducial run in a taller box of vertical size $Lz = 8H_0$ (NSTRMRI4e). We have also investigated the effect of varying the radial box size.

\subsection{Set-up}
\label{BOXCOOLING_Setup}

The simulation parameters and set-up follows the template described in Section 2: the resolution was $128\times128\times196$ and the box size was $4H_0\times4H_0\times6H_0$. It was run for 200 orbits ($1257\, \Omega^{-1}$). In order to isolate the behavior of the MRI in vertically stratified boxes with a perfect gas equation of state, we omit explicit cooling and diffusion coefficients. We employ a mass source term to replace mass lost through the vertical boundaries. The mass loss per orbit is is around $3\%$ of the initial box mass $M_0$. One disadvantage of this approach, of course, is that the thermal equilibrium will be influenced by the size of the box and the vertical boundary conditions. Employing taller boxes does not alleviate the issue as the disc will always expand until advective cooling across the vertical boundaries balances viscous heating by the MRI at the mid-plane no matter how tall the box is. To check this explicitly, we have repeated this simulation in a taller box of $L_z = 8H_0$ (keeping everything else, including the number of grid cells per scale-height, identical to the $L_z = 6H_0$ run). As expected the mass loss per orbit is only marginally smaller than that measured in the shorter box run, and appears to converge with that measured in the shorter box run towards the end of the simulation.
 
\subsection{Time evolution of averaged quantities}
\label{BOXCOOLING_TimeEvolutionOfAveragedQuantities}
We first present a number of classical box-averaged diagnostics to show that the MRI turbulence we generate is in line with previous results in the literature and is behaving as expected. After initialization, there is exponential growth of the volume averaged energies (corresponding to the linear phase of the MRI) over the first few orbits, followed by non-linear saturation. Time-averaged stresses are given in table \ref{TABLE_MRISimulationsWithoutCoolingDependenceOnBoxSize}. The volume-averaged pressure is correlated with $\langle \alpha \rangle$, and the ratio of magnetic stress to Reynolds stress is $\sim4.5$. 

The thermal energy balance is determined by turbulent dissipation of heat due to the MRI and cooling due to advection of thermal energy across the vertical boundaries. The horizontally-averaged mid-plane scale height $\langle H \rangle_{xy}(z=0,t) \equiv \sqrt{\gamma \langle T(z=0,t)\rangle_{xy}} /\Omega$ is a good preliminary diagnostic of the variation in `disc thickness' over the course of the simulation. After initialization $\langle H \rangle_{xy}$ increases rapidly from  $\langle H \rangle_{xy}=H_0$ reaching $\langle H\rangle_{xy} \sim 1.5H_0$ at non-linear saturation (around orbit 7), before leveling out and fluctuating just under $1.5H_0$ for the remainder of the simulation. This is well below the vertical box semi-size of $L_z / 2 = 3H_0$, thus the numerical domain remains large enough to describe the disc's natural thickness. This is quite different to the `tent-like' profiles obtained by \cite{bodo2012} and \cite{bodo2014} in which the disc heated up to the point that $H\gg L_z$.  For a more quantitative estimate of the effect of box cooling we also calculated the wind cooling time (cf. Equation \ref{EQUN_WindCoolingRate}) and found this to be $\tau_\text{w} \sim 112\, \Omega^{-1}$ (time-averaged between orbit 125 and orbit 195) compared to the value of $\tau_\text{w} \sim 150 \,\Omega^{-1}$ measured in the box-cooled simulation MRI-S1 of \cite{riols2018}.

\begin{figure}
\centering
\includegraphics[scale=0.33]{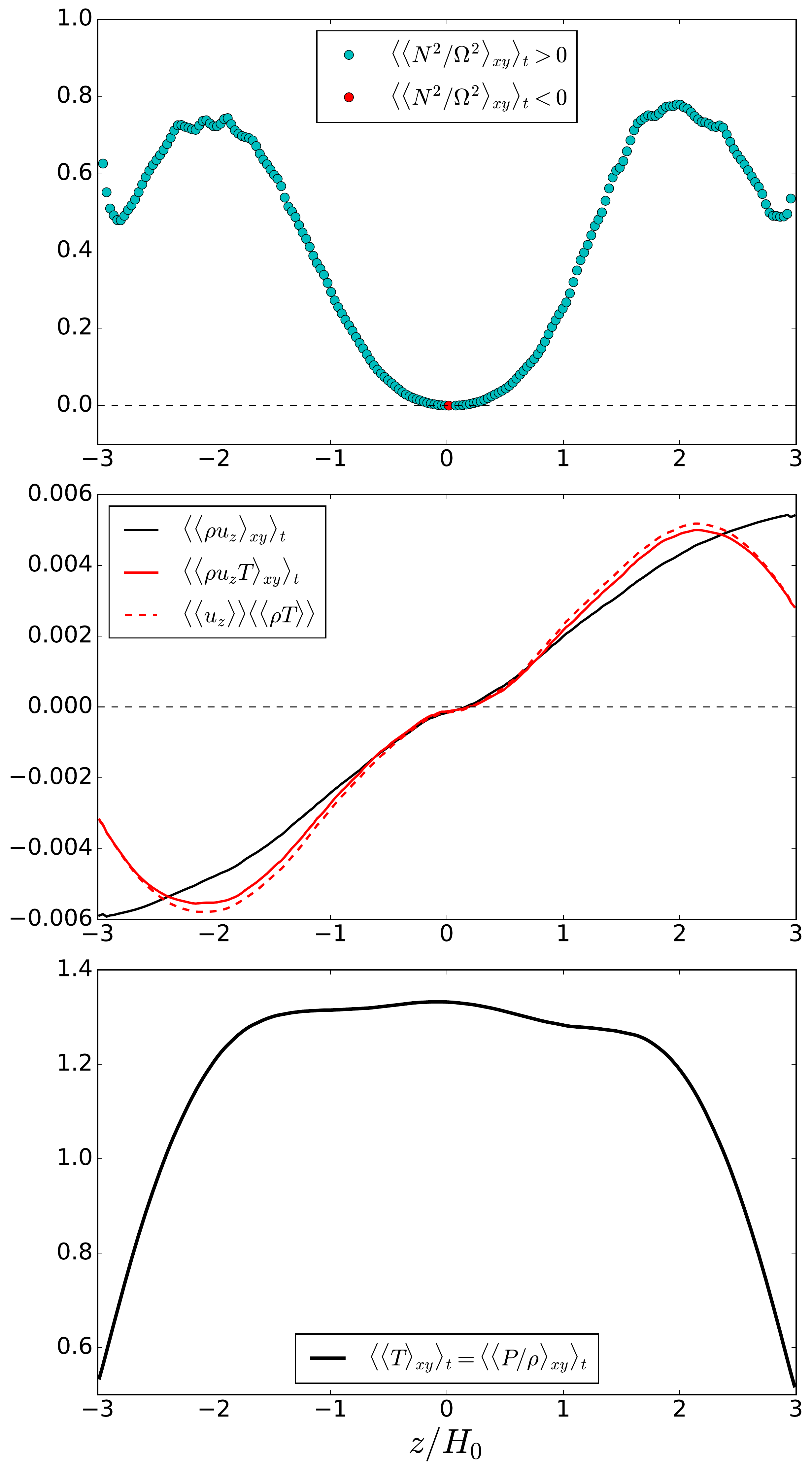}
\caption{Horizontal-and-time-averaged vertical disc profiles from the fiducial inviscid box-cooled vertically stratified MRI simulation without explicit cooling. Top: (square) of the buoyancy frequency. Middle: vertical mass (in black) and heat (in red) fluxes. Bottom: temperature profile. The time-averages are taken from orbit 40 to orbit 196.}
\label{FIGURE_FiducialSimsVerticalProfiles}
\end{figure}

\subsection{Vertical structure of the disc}
\label{BOXCOOLING_VerticalStructure}
In Figure \ref{FIGURE_FiducialSimsVerticalProfiles} we plot the horizontal- and time-averaged vertical profiles of various quantities. A surprising result here is the large outward heat flux. First note the vertical profile of the buoyancy frequency $\langle \langle N_B^2 \rangle_{xy} \rangle_t$ (top panel), which is positive everywhere, except exactly at the mid-plane where the disc is marginally stable. The positivity of $N_B^2$ throughout the domain precludes convection, and indeed the flow field (i.e. $u_z$ in the $xz$-plane) appears to be dominated by MRI turbulence: we do not observe any obvious vertical structure or plumes characteristic of convection. The vertical specific kinetic energy is distributed isotropically in the $(k_x,k_z)$ plane (see upper-left-hand panel of Figure \ref{FIGURE_2DPowerSpectraCoolingPrescriptionsComparison}). Even in the limit of efficient convection, the system would only be brought to marginality, not to $N^2_B>0$ as we see. 

Despite the absence of convection there is a significant vertical heat flux, as shown in the middle panel (solid red curve). The heat flux peaks at a value of around $F_{\text{heat}} \sim \pm\, 5\times10^{-3}$ at around $\pm\, 2H_0$ on either side of the mid-plane, significantly larger than the heat flux of around $5\times10^{-5}$ carried by convection in our hydrodynamic simulations (cf. Figure 2 of \cite{heldlatter2018}).

Given that the disc is convectively stable it is tempting to ascribe this large outward heat flux with transport due to MRI turbulence. It is worrying, however, that the heat flux closely tracks the outward mass flux (solid black curve). This seems to suggest that this outward transport of heat is not due to the MRI per se, but due to a mean flow that results from box-cooling. In short we are adding mass in by hand and it is leaving through the vertical boundaries, setting up the aforementioned mean vertical flow. This is confirmed by plotting the quantity $\langle \langle u_z \rangle \rangle \langle \langle \rho T \rangle \rangle$ (dashed red curve): the difference between the mean vertical heat transport $\langle \langle \rho u_z T \rangle \rangle$ and this quantity should be more representative of the actual turbulent heat transport of heat. We find that the two curves nearly overlap, however, which seems to suggest that there is actually very little vertical transport of heat due to MRI turbulence: instead the outward heat flux is almost entirely due to the mean vertical flow which itself is an artifact of box-cooling.

\begin{figure}
\centering
\captionsetup[subfigure]{labelformat=empty}
\subfloat[]{\includegraphics[scale=0.22]{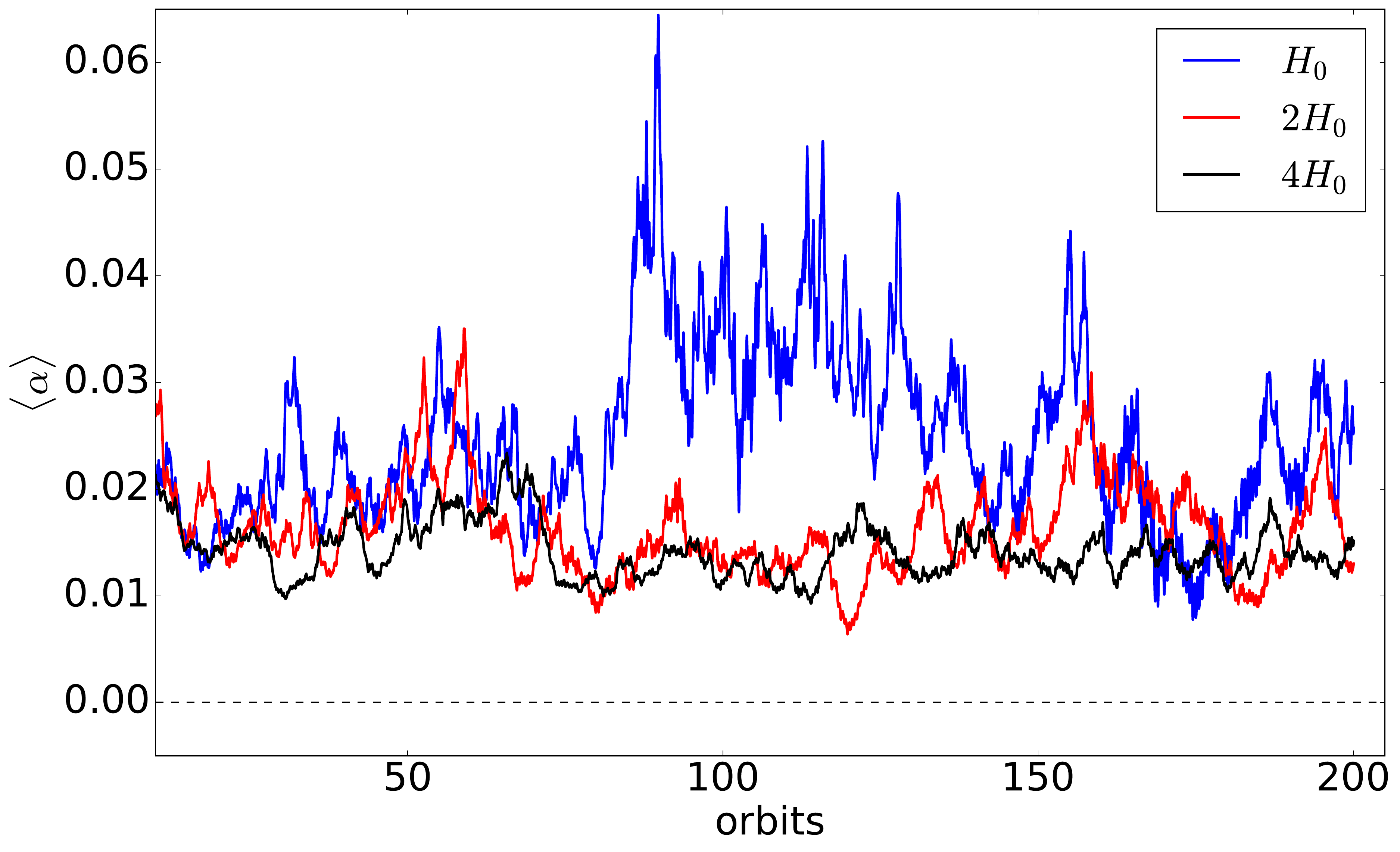}}
\caption{Plot of time-evolution of volume-averaged $\alpha$ for vertically stratified MRI simulations (without explicit cooling or diffusion coefficients) at three different radial box sizes: $L_x=H_0$ (blue), $L_x=2H_0$ (red), and $L_x=4H_0$ (black). All three simulations were run with a resolution of $32$ cells / $H_0$ in each direction.}
\label{FIGURE_ConvergenceStudies}
\end{figure}

The density (not shown) drops monotonically from the mid-plane and is about two orders of magnitude smaller at the vertical boundaries than at the mid-plane. We do not observe the `tent-like'  temperature and flat density profiles reported in \cite{bodo2012}, both of which are indicative that the disc has filled the box.
The vertical stress profiles for the time- and horizontally-averaged Reynolds stress $\langle \langle R_{xy}\rangle_{xy}\rangle_t$ and magnetic stress $\langle \langle M_{xy}\rangle_{xy}\rangle_t$ (not shown) are relatively constant within  $\pm 1.5H_0$ of the mid-plane. Both profiles drop off rapidly  beyond $\pm1.5H_0$ and are only about one tenth of their peak values at the boundaries. 

Finally, we have also investigated the effect of varying the radial box size. In Figure \ref{FIGURE_ConvergenceStudies} we plot the time-evolution over 200 orbits of $\langle \alpha \rangle$ for the three different radial box sizes. The time-averaged value of $\alpha$ \textit{increases} as the radial box size \textit{decreases} (see Table \ref{TABLE_MRISimulationsWithoutCoolingDependenceOnBoxSize}). Fluctuations in $\langle \alpha \rangle$ also increase markedly as the radial box size is decreased. The difference is particularly noticeable when comparing the $L_x=H_0$ simulation with the other two simulations. The buoyancy frequency is positive everywhere in all three simulations, but both the vertical heat and mass fluxes increase as the radial box size is reduced, which is unsurprising given that fluid elements are increasingly constrained to move in the vertical direction. We conclude that the non-linear MRI is almost certainly not converged in boxes of radial size $L_x = H_0$. 
In particular, narrow boxes seem to give rise to large outbursts that are not observed in wider boxes. Therefore it would be illuminating to investigate whether the MRI/radiative cycles observed in the simulations of \cite{hirose2014} (which were carried out in boxes of radial size $L_x = H_0$) are less pronounced or even absent in wider boxes.

\section{Simulations with uniform explicit cooling}
\label{UNIFORMCOOLING}
The inviscid (or ideal) simulations discussed in Appendix \ref{BOXCOOLING} are \textit{box-cooled}, in the sense that cooling is facilitated by advection of thermal energy across the vertical boundaries only. As demonstrated in Appendix \ref{BOXCOOLING_VerticalStructure} box-cooling results in a large outward heat flux (irrespective of vertical box size) that has nothing to do with turbulent transport. Thus we have also run a series of simulations with identical resolution ($128\times128\times196$ or $32/H_0$), box size size ($4H_0\times4H_0\times6H_0$), and initial conditions to our fiducial box-cooled simulation, but \textit{including} explicit cooling via a linear cooling prescription (see Section \ref{METHODS_GoverningEquations}). We investigated cooling timescales between $\tau_\text{c}=5$ and 30 orbits. In contrast to the simulations discussed in Section \ref{HEIGHTDEPCOOLING}, cooling is employed \textit{uniformly} everywhere in the box in these runs, and no explicit diffusion coefficients are used. The simulations were run for 100-200 orbits. 

\begin{figure}
\centering
\includegraphics[scale=0.17]{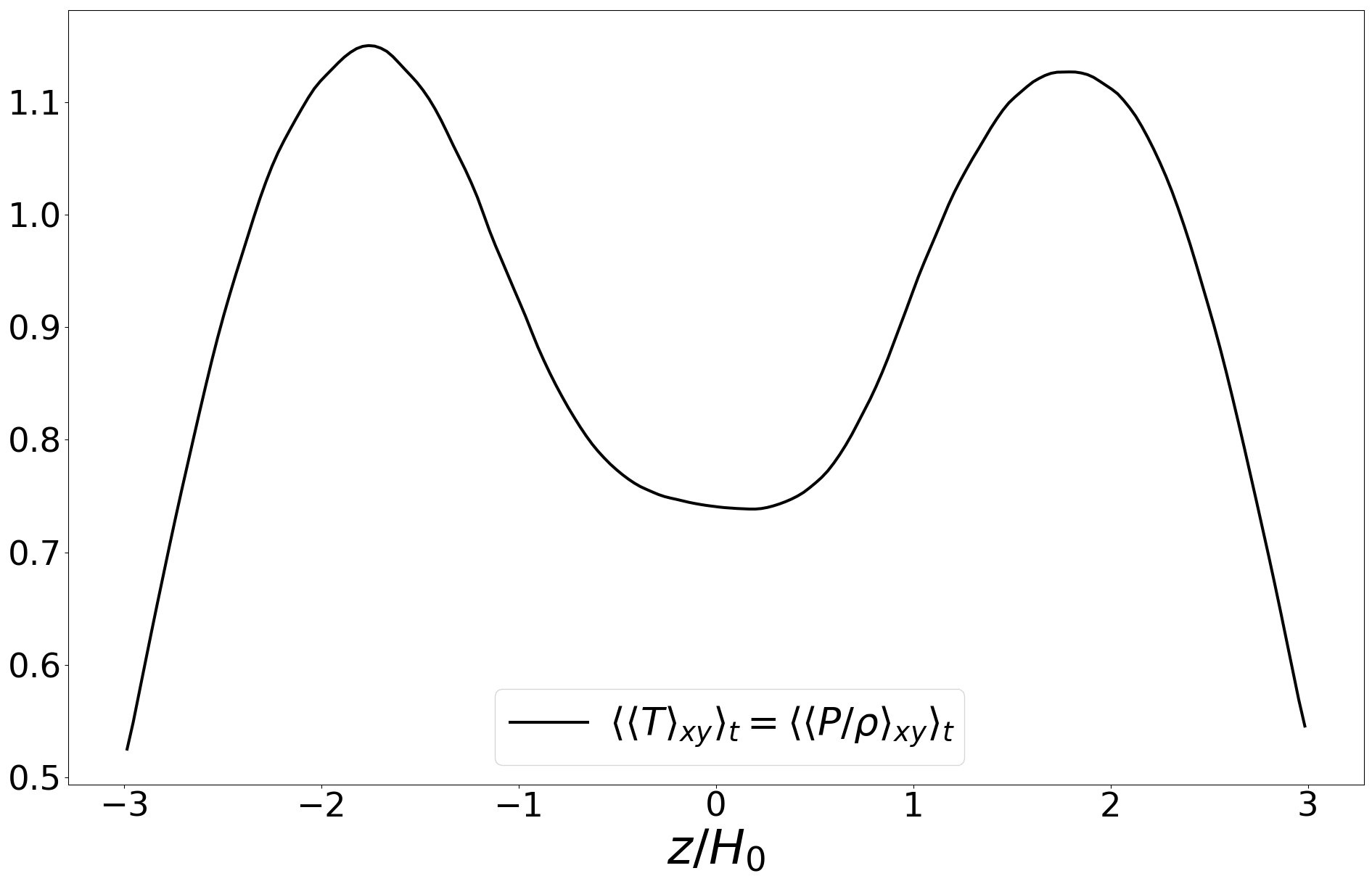}
\caption{Horizontal- and time-averaged vertical temperature profile from a simulation with a uniform explicit cooling time-scale of $\tau_c = 17$ orbits. The mid-plane is in thermal equilibrium and explicit cooling dominates box cooling, but there is a marked dip in the mid-plane temperature, a pathology characteristic of runs with uniform cooling.}
\label{FIGURE_VerticalTempProfileTauc17Orb}
\end{figure}

We have found that such uniformly cooled runs exhibit their own challenges and pathologies. For thermal equilibrium to be maintained in the non-linear phase, total cooling (explicit and box-cooling) must balance viscous heating. If the cooling is too strong (i.e. $\tau_c$ is too small) the cooling rate will exceed the MRI heating rate and the disc will contract, but if it is too weak (i.e. $\tau_c$ is too large) the disc will expand until thermal equilibrium is reached by advection of fluid across the vertical boundaries (box-cooling). The latter quantity is measured by the wind cooling timescale $\tau_\text{w}$ (see Equation \ref{EQUN_WindCoolingRate}): thus for explicit cooling to dominate we require that $\tau_c < \tau_\text{w}$.  

We have found three different outcomes across the range of cooling timescales investigated. When $\tau_c \leq 13$ orbits, cooling is strongly dominated by explicit cooling (e.g. in the $\tau_c = 13$ orbits run $\tau_\text{w} \sim 104.2\,\Omega^{-1}$ [time-averaged between orbit 125 and 200] compared to $\tau_c = 81.7\,\Omega^{-1}$), but we observe a large dip in the mid-plane temperature, which only worsens with time. For example, in the $\tau_c = 13$ orbits simulation the (horizontal- and time-averaged) mid-plane temperature $\langle \langle T(z=0) \rangle_{xy} \rangle$ is around $0.9$ between orbits 40 and 100, but drops to $0.4$ between orbits 125 and 200. In the $\tau_c = 5$ orbits run, the mid-plane temperature reaches as low as $0.2$ within the first 100 orbits after initialization. In all these runs the drop in mid-plane temperature is accompanied by a marked drop in the mid-plane stresses. It appears that MRI heating cannot balance the explicit cooling at the mid-plane and the disc contracts. Thus while explicit cooling dominates box-cooling in these runs, they are \textit{not} in thermal equilibrium (at least at the mid-plane). 

At more moderate cooling timescales of $\tau_c = 15$ and $17$ orbits explicit cooling still dominates over wind cooling (e.g. $\tau_c = 106.8\,\Omega^{-1}$ compared to $\tau_\text{w} \sim 117.0\,\Omega^{-1}$ [time-averaged between orbits 125 and 175] in the $\tau_c = 17$ orbits run). We still get a moderate dip in the mid-plane temperature (see Figure \ref{FIGURE_VerticalTempProfileTauc17Orb}), but it appears to be steady over the 200 orbits for which the simulations were run. Therefore these runs are both explicit-cooling-dominated, \textit{and} in thermal equilibrium, and represent a kind of best-case scenario for uniformly cooled simulations.

Finally for $\tau_c \gtrsim 18$ orbits, box-cooling begins to dominate (i.e. $\tau_\text{w} \lesssim \tau_c$). By $\tau_c = 30$ the vertical disc structure is comparable to that found in our box-cooled simulations (cf. Figure \ref{FIGURE_FiducialSimsVerticalProfiles}), though this is not surprising given that the simulation is strongly dominated by box-cooling, with $\tau_\text{w} \sim 95\, \Omega^{-1}$ (time-averaged between orbit 150 and 200) compared to $\tau_c \sim 188\,\Omega^{-1}$.

Note that despite the pathologies exhibited in the vicinity of the mid-plane by the uniformly cooled runs, the volume-averaged total stress is comparable to that found in our fiducial box-cooled run, and the disc is convectively stable (i.e. $\langle \langle N_B^2 \rangle_{xy} \rangle_t > 0$ for all $|z| > 0$) in all of our uniformly cooled runs. Outside of the mid-plane region the heat flux is directed outwards and is of similar magnitude to that measured in our box-cooled run.

\section{Tables of Simulations}
\label{APPENDIX_TablesOfSimulations}

\begin{table*}
\centering
\caption{Vertically stratified MHD simulations with height-dependent cooling and explicit resistivity. All simulations were run in a box of size $4H_0\times4H_0\times6H_0$ and a resolution of $128\times128\times196$ (32 cells per $H_0$) except NSTRMC44f1HR (which was run with a resolution of $256\times256\times392$) and NSTRMC44e1TB (which was run in a taller box of vertical extent $L_z = 8H_0$). The explicit resistivity is denoted by $\eta$, $\tau_\text{c}$ denotes the cooling timescale (in orbits), Rm$\equiv H_0 c_{s0} / \eta$ is the magnetic Reynolds number, $\Delta T$ denotes the duration of the simulation (in orbits), max$(F_\text{heat})$ denotes the maximum value of the horizontally- and time-averaged vertical heat flux within $\pm H_0$, and min$(N_B^2)$ denotes the minimum value of the horizontally- and time-averaged square of the buoyancy frequency. Time-averages were typically taken over the last 80 orbits of a simulation.}
\label{TABLE_MRISimulationsWithCooling}
	\begin{tabular}{lccccccr}
		\hline
		Run	& $\tau_\text{c}$ & $\eta$ & Rm & $\Delta T$ &Result &  max$(F_\text{heat})$ & min$(N_B^2)$ \\ 
		\hline
		NSTRMC44e1      & $10$  &  $5\times10^{-3}$    & $200$ & 200  & strong MRI/convective cycles & 0.0044 & $-0.09$\\
		NSTRMC44e1TB & $10$  &  $5\times10^{-3}$    & $200$ & 200  & strong MRI/convective cycles & 0.0063 & $-0.08$\\

		NSTRMC44e2      & $5$    &  $5\times10^{-3}$    & $200$ & 100 & strong MRI/convective cycles & 0.0062 & $-0.14$\\
		\hline 
		NSTRMC44a1      & $10$  &  $1\times10^{-3}$    & $10^3$ & 100  & strong MRI/convective cycles & 0.0012 & $-0.04$\\
		NSTRMC44a2      & $5$    &  $1\times10^{-3}$    & $10^3$& 110 & strong MRI/convective cycles & 0.0018 & $-0.17$\\
		\hline
		NSTRMC44b1      & $10$  &  $5\times10^{-4}$    & $2\times10^3$ & 100  & strong MRI/convective cycles & 0.0004 & $-0.07$\\
		NSTRMC44b2      & $5$    &  $5\times10^{-4}$    & $2\times10^3$ & 100  & strong MRI/convective cycles & 0.0017 & $-0.17$\\
		NSTRMC44b4      & $3$    &  $5\times10^{-4}$    & $2\times10^3$ & 100  & strong MRI/convective cycles & 0.0003 & $-0.25$ \\
		\hline
		NSTRMC44g1      & $10$    &  $2.5\times10^{-4}$ & $4\times10^3$ & 100  & straight MRI                 & 0.0007 &  $-0.12$\\
		NSTRMC44g4      & $3$    &  $2.5\times10^{-4}$ & $4\times10^3$ & 100  & straight MRI                 & 0.0003 &  $-0.30$\\
		NSTRMC44g5      & $2$    &  $2.5\times10^{-4}$ & $4\times10^3$ & 200  & weak MRI/convective cycles & 0.0005 &  $-0.43$\\
		NSTRMC44g6      & $1$    &  $2.5\times10^{-4}$ & $4\times10^3$ & 200  & weak MRI/convective cycles & 0.0006 &  $-0.51$\\
		\hline
		NSTRMC44c1     & $10$  &  $1\times10^{-4}$    & $10^4$ & 100  & straight MRI                  & 0.0013 &  $-0.14$\\
		NSTRMC44c2     & $5$    &  $1\times10^{-4}$    & $10^4$ & 80    & straight MRI                  & 0.0020 &  $-0.24$\\
		NSTRMC44c4     & $3$    &  $1\times10^{-4}$    & $10^4$ & 320  & straight MRI                  & 0.0022 &  $-0.34$\\
		NSTRMC44c5     & $2$    &  $1\times10^{-4}$    & $10^4$ & 100  & straight MRI                  & 0.0012 &  $-0.43$\\
		NSTRMC44c6     & $1$    &  $1\times10^{-4}$    & $10^4$ & 300  & weak MRI/convective cycles  & 0.0011 &  $-0.59$\\
		NSTRMC44c7     & $0.5$ &  $1\times10^{-4}$    & $10^4$ & 100  & disc disrupted                & 0.0007&  $-0.72$\\
		\hline
		NSTRMC44d1     & $10$  &  $5\times10^{-5}$    & $2\times10^4$ & 100  & straight MRI                  & 0.0013 & $-0.16$\\
		NSTRMC44d2     & $5$    &  $5\times10^{-5}$    & $2\times10^4$ & 100  & straight MRI                  & 0.0020 & $-0.24$\\
		\hline
		NSTRMC44f1      & $10$  &  $1\times10^{-5}$    & $10^{5}$ & 100  & straight MRI                  & 0.0018 & $-0.15$\\	
		NSTRMC44f1HR & $10$  &  $1\times10^{-5}$    & $10^{5}$ & 75    & straight MRI                  & 0.0029 & $-0.21$\\	
		NSTRMC44f2      & $5$  &  $1\times10^{-5}$    & $10^{5}$ & 100  & straight MRI                  & 0.0049 & $-0.25$\\	
		\hline
		NSTRMC44e1NoCool & $\infty$ &  $5\times10^{-3}$ &  $200$ & 100  & No MRI; no convection & 0.00014 & $\sim 0$\\
		NSTRMC44b1NoCool & $\infty$ &  $5\times10^{-4}$ & $2\times10^3$ &100  & MRI; no convection & 0.00137 & $\sim 0$\\
		\hline	
	\end{tabular}
\end{table*}

\begin{table*}
\centering
\caption{Vertically stratified MHD simulations with height-dependent cooling and explicit resistivity: effect of changing radial box size. The column headers have the same meaning as in Table \ref{TABLE_MRISimulationsWithCooling}. The number of bursts refers only to those within the first 100 orbits of initialization.}
\label{TABLE_MRISimulationsWithCooling2}
	\begin{tabular}{lccccccccr}
		\hline
		Run	&  Radial Box Size & Resolution & $\tau_\text{c}$ & $\eta$ & Rm & $\Delta T$ &Result &  Number of bursts \\ 
		\hline
		NSTRMC44e1(1H) & $H_0$    & $32\times128\times196$   &  $10$ & $5\times10^{-3}$  & 200 & 100  & MRI/convective cycles  & 1 \\
		NSTRMC44e1(2H) & $2H_0$  & $64\times128\times196$   & $10$  & $5\times10^{-3}$  & 200 & 100  & MRI/convective cycles   & 2 \\
		NSTRMC44e1 & $4H_0$  & $128\times128\times196$ & $10$  & $5\times10^{-3}$  & 200 & 200  & MRI/convective cycles & 3\\
		\hline
	\end{tabular}
\end{table*}

\begin{table*}
\centering
\caption{Vertically stratified ideal MHD simulations without explicit cooling (box-cooled simulations). Time-averages and maxima/minima were taken from orbit 50 to orbit 200, except for NSTRMRI4d for which averages were taken from orbit 40 to orbit 100.}
\label{TABLE_MRISimulationsWithoutCoolingDependenceOnBoxSize}
	\begin{tabular}{lccccccccr}
		\hline
		Run	& $L_x$ & $L_z$ & Resolution & Grid cells / $H_0$ &   $\langle\langle\alpha\rangle\rangle$ & $\langle\langle R_{xy}\rangle\rangle$ & $\langle\langle M_{xy}\rangle\rangle$ & $\text{max}(\langle\alpha\rangle)$  & $\text{min}(\langle\alpha\rangle)$\\
		\hline
		NSTRMRIf3a    & $H_0$    & $6H_0$  & $32\times128\times196$ & $32/H_0$ & $0.0262$ &  $0.0026$  & $0.0123$ & $0.0645$ & $0.0078$\\
		NSTRMRI5       & $2H_0$  & $6H_0$ & $64\times128\times196$ & $32/H_0$ & $0.0161$ & $0.0016$ & $0.0070$ & $0.0350$ & $0.0064$\\
		NSTRMRI4c     & $4H_0$  & $6H_0$ & $128\times128\times196$ & $32/H_0$ & $0.0140$ & $0.0014$ & $0.0061$ & $0.0234$ & $0.0093$ \\
		NSTRMRI4d     & $4H_0$  & $6H_0$ & $256\times256\times392$ & $64/H_0$ & $0.0132$ & $0.0013$ & $0.0058$ & $0.0115$ & $0.0161$ \\
		NSTRMRI4e   & $4H_0$  & $8H_0$ & $128\times128\times256$ & $32/H_0$ & $0.0135$ & $0.0016$ & $0.0071$ & $0.0191$ & $0.0091$ \\
		\hline
	\end{tabular}
\end{table*}

\bsp	% typesetting comment
\label{lastpage}
\end{document}